\documentclass[11pt]{article}
\pdfoutput=1
\usepackage{jcapmod}

\usepackage{amsfonts, amsmath}
\usepackage{booktabs}
\usepackage[english]{babel}
\usepackage{colortbl}
\usepackage{graphicx}
\usepackage{hyperref}
\usepackage{framed}
\usepackage{simplewick}
\usepackage{siunitx}
\usepackage{tikz}
\usepackage{listings}

\usepackage[T1]{fontenc}
\usepackage{lmodern}
\usepackage{geometry}
\usepackage{amsmath, amssymb}
\usepackage{physics}
\usepackage{pifont}
\usepackage{siunitx}
\usepackage{hyperref}
\usepackage{marginnote}
\usepackage{xcolor}
\usepackage{listings}
\usepackage{enumitem}

\definecolor{CiteBlue}{RGB}{45,52,151}
\hypersetup{colorlinks,
  linkcolor=blue!60!black,
  citecolor=blue!50!black,
  urlcolor=cyan!60!black}

\lstdefinestyle{fortran}{
  language={[90]Fortran},
  basicstyle=\ttfamily\small,
  keywordstyle=\color{blue!70!black},
  commentstyle=\color{green!45!black}\itshape,
  stringstyle=\color{red!60!black},
  numbers=left, numberstyle=\tiny, stepnumber=1,
  numbersep=5pt,
  breaklines=true
}

\RequirePackage{color}

\usepackage{colortbl}
\usepackage[normalem]{ulem}
\usepackage{comment}

\definecolor{rp}{cmyk}{0.2, 1, 0.6, 0}
\definecolor{green2}{cmyk}{0, 1, 0.5, 0}
\definecolor{lightgreen}{cmyk}{0.2, 0, 0.2, 0.2}
\definecolor{lightgray}{cmyk}{0.1,0.2,0,0.1}
\definecolor{lightgray2}{cmyk}{0.4,0.4,0,0.8}
\definecolor{black}{cmyk}{1.0,1.0,1.0,1.0}

\allowdisplaybreaks[1]

\usepackage{colortbl}
\definecolor{lightgreen}{cmyk}{0.2, 0, 0.2, 0.2}
\definecolor{lightgray}{cmyk}{0.1,0.2,0,0.1}
\definecolor{lightgray2}{cmyk}{0.1,0.1,0,0.1}

\definecolor{NewRed}{RGB}{200,37,6}
\definecolor{NewOrange}{RGB}{222,106,16}
\definecolor{NewGreen}{RGB}{0,136,43}

\makeatletter
\newlength{\apb@width}
\newcommand{\autoparbox}[2][c]{\settowidth{\apb@width}{#2}\parbox[#1]{\apb@width}{#2}}

\makeatother

\numberwithin{equation}{section}

\def\beq{\begin{equation}}
\def\eeq{\end{equation}}
\def\bea{\begin{eqnarray}}
\def\eea{\end{eqnarray}}

\def\0{{\boldsymbol 0}}

\DeclareRobustCommand{\SkipTocEntry}[4]{}

\begin{document}

\begin{titlepage}

\title{\centering Cosmological gravitational particle production in multifield inflation}

\author[a,b]{Edward W. Kolb,}
\author[a]{Sarunas Verner,}
\author[a]{and Jingyuan Wang}

\affiliation[a]{Kavli Institute for Cosmological Physics, The University of Chicago, 5640 South Ellis Avenue, Chicago, IL 60637, U.S.A.}
\affiliation[b]{Enrico Fermi Institute, The University of Chicago, 5640 South Ellis Avenue, Chicago, IL 60637, U.S.A.}
\emailAdd{Rocky.Kolb@uchicago.edu}
\emailAdd{verner@uchicago.edu}
\emailAdd{jywang7@uchicago.edu}

\abstract{
We study cosmological gravitational particle production (CGPP) of dark matter in two-field inflationary backgrounds with both flat and curved field-space geometries. As a concrete realization of broader multifield mechanisms, we adopt a Starobinsky\,+\,quadratic potential and construct benchmark scenarios that interpolate between the flat field-space limit and the sidetracked attractor on a hyperbolic field space, and we compute the production spectrum of a gravitationally coupled spectator scalar for both minimal ($\xi = 0$) and conformal ($\xi = 1/6$) coupling to the Ricci scalar. We show that negative field-space curvature can strongly enhance the post-inflationary oscillations of the Ricci scalar, leading to up to an order of magnitude enhancement of the CGPP number density relative to the flat field-space limit, particularly for minimal coupling. For the sidetracked attractor, this enhanced production competes with a reduced inflationary energy scale, leading to a nontrivial dependence of the relic abundance on model parameters. We derive the relic abundance as a function of spectator mass and reheating temperature, and identify the viable parameter space for each benchmark. The conformal case $\xi = 1/6$, whose scalar mode equation is structurally analogous to that of a massive Dirac fermion, is much less constrained by isocurvature and provides a minimal scenario for purely gravitational dark matter production in multifield inflation.
}


\maketitle

\end{titlepage}

\setcounter{page}{1} 

\section{Introduction}
\label{sec:intro}
Cosmic inflation remains the leading paradigm for generating the initial conditions of the observable universe, producing a nearly scale-invariant, Gaussian spectrum of primordial perturbations that seeds the large scale structure of the cosmos. Precision measurements of the cosmic microwave background (CMB) by Planck~\cite{Planck:2018jri,Planck:2018vyg}, the Atacama Cosmology Telescope (ACT)~\cite{AtacamaCosmologyTelescope:2025blo,AtacamaCosmologyTelescope:2025nti}, and the South Pole Telescope (SPT)~\cite{SPT-3G:2024atg,SPT-3G:2025bzu} confirm a red-tilted scalar spectrum ($n_s<1$) at high significance and, together with B-mode polarization data from BICEP/Keck~\cite{BICEP:2021xfz}, constrain the tensor-to-scalar ratio to $r<0.036$ at 95\% CL. These increasingly precise constraints place some previously favored single-field models under growing pressure~\cite{Ellis:2025zrf,Martin:2024qnn}. In this evolving landscape, multifield models of inflation, which arise generically in ultraviolet-motivated frameworks such as string compactifications and supergravity~\cite{Baumann:2014nda,Cicoli:2023opf}, have received growing attention both for their richer phenomenology and for their ability to populate regions of the $(n_s,r)$ plane that are difficult to realize in simple single-field constructions.

Independent of the inflationary model, the expansion of the universe during and after inflation produces particles through the purely gravitational mechanism of \emph{cosmological gravitational particle production} (CGPP)~\cite{Parker:1969au,Zeldovich:1971mw,Ford:1986sy,Chung:1998zb,Chung:1998ua,Kolb:2023ydq}. Any field whose dynamics break conformal invariance, through a mass term, nonminimal coupling to gravity, or nontrivial spin structure, can be produced by the time-varying spacetime geometry, provided that the expansion is sufficiently nonadiabatic~\cite{Birrell:1982ix,Parker:2009uva}. This makes CGPP a uniquely minimal mechanism for generating dark matter: it requires no portal couplings and no thermal contact with the Standard Model~\cite{Chung:1998zb,Chung:1998ua,Chung:2001cb}. The resulting ``WIMPzilla'' or superheavy dark matter scenario~\cite{Kolb:1998ki} has been developed extensively for single-field inflationary backgrounds, with detailed calculations of the production spectra for scalars~\cite{Chung:1998zb,Chung:2001cb,Ema:2018ucl,Ema:2019yrd,Ling:2021zlj,Garcia:2022vwm,Garcia:2023awt,Garcia:2023qab,Jenks:2024fiu,Racco:2024aac,Verner:2024agh}, fermions~\cite{Chung:2011ck,Ema:2019yrd}, vectors~\cite{Graham:2015rva,Ahmed:2020fhc,Kolb:2020fwh,Capanelli:2024rlk}, and massive spin-2 particles~\cite{Kolb:2023dzp, Verner:2025ise}. The quantum-interference structure of the post-inflationary spectra has been analyzed in Ref.~\cite{Basso:2022tpd}, and a comprehensive review with a more complete reference list can be found in Ref.~\cite{Kolb:2023ydq}.

Despite this progress, the overwhelming majority of CGPP calculations have been performed on single-field inflationary backgrounds. This is a significant limitation, because the CGPP spectrum is determined not by the inflationary potential alone, but by the full time evolution of the scale factor $a(\eta)$ and the Ricci scalar $R(\eta)$ through the spectator field mode equation, and these background quantities can differ qualitatively between single-field and multifield models. In particular, the post-inflationary oscillation phase, during which the inflaton fields oscillate about the potential minimum and the Ricci scalar undergoes repeated zero crossings that drive bursts of particle production~\cite{Chung:1998zb,Basso:2022tpd}, is generically modified in multifield models~\cite{Watanabe:2015eia}: the presence of a second oscillating field introduces additional frequencies and beating patterns into $R(\eta)$, while curved field-space geometries, which arise naturally in supergravity and string compactifications~\cite{Kallosh:2013hoa,Ferrara:2013rsa,Galante:2014ifa,Cicoli:2018kdo}, can couple the oscillation modes through field-space Christoffel symbols and strongly enhance the kinetic energy through nontrivial metric factors. These effects have no analogue in single-field inflation and can modify the CGPP spectrum across both the infrared and ultraviolet regimes.

An important step toward understanding CGPP in multifield inflation was taken in Ref.~\cite{Kolb:2022eyn}, which studied gravitational particle production in rapid-turn multifield models motivated in part by the de~Sitter swampland conjecture. That work demonstrated that sustained turning in field space can modify the production spectrum and identified viable dark matter parameter space consistent with relic abundance and isocurvature constraints. In the present work, we take a complementary approach: rather than studying a single multifield model, we construct a comprehensive set of benchmarks that systematically vary both the initial conditions in field space and the field-space curvature, allowing us to disentangle the separate effects of multifield dynamics and field-space geometry on CGPP.

Our framework is based on the Starobinsky\,+\,quadratic potential, $V = V_{\rm S}(\phi_S) + \frac{1}{2}m_Q^2\phi_Q^2 \,,$ where $V_{\rm S}$ is the Starobinsky potential~\cite{Starobinsky:1980te}. This potential is placed on a two-dimensional field space with either flat or hyperbolic geometry, the latter parametrized by a curvature $L_{\rm fs}$. On a flat field space, we construct a ten point scan (SQ$_1$--SQ$_{10}$) that interpolates between the pure quadratic and pure Starobinsky limits, probing how the relative weighting of the two sectors modifies the post-inflationary oscillation pattern and hence the CGPP spectrum. On a hyperbolic field space, we study the phenomenon of \emph{sidetracked inflation}~\cite{Renaux-Petel:2015mga,Garcia-Saenz:2018ifx,Garcia-Saenz:2018vqf}, an inflationary attractor with sustained turning induced by the geometrical destabilization of the inflaton valley~\cite{Renaux-Petel:2015mga}. We scan the parameter ratio $L_{\rm fs}/L_{\rm crit}$\footnote{In sidetracked inflation, for $L_{\rm fs} < L_{\rm crit}$ the valley floor is tachyonically unstable, see Eq.~\eqref{eq:Lcrit}. } from the flat limit through the stable regime and into the sidetracked regime, at two mass ratios, producing benchmarks that span from perturbative corrections to Starobinsky inflation through a CMB excluded band near the destabilization threshold to a genuinely distinct attractor with $r \sim 10^{-7}$. The hyperbolic geometry connects directly to the $\alpha$-attractor framework arising in $\mathcal N = 1$ supergravity~\cite{Kallosh:2013hoa,Ferrara:2013rsa,Kallosh:2013yoa,Galante:2014ifa,Ellis:2013nxa, Ellis:2019bmm, Ellis:2021kad, Ellis:2025zrf}, providing a concrete UV-motivated context for the field-space curvature effects we study.

For each benchmark, we compute the CGPP spectrum of a spectator scalar $\chi$ with both minimal ($\xi = 0$) and conformal ($\xi = 1/6$) coupling to the Ricci scalar. The conformal case is of particular interest because the mode equation for a conformally coupled massive scalar is structurally analogous to that of a massive Dirac fermion minimally coupled to gravity, sharing the same dominant time-dependent frequency squared $k^2 + m^2 a^2$~\cite{Kolb:2023ydq}. It therefore provides a useful proxy for fermionic dark matter production. Furthermore, conformal coupling is much less constrained by isocurvature than the minimally coupled light spectator case, which is typically disfavored when $\chi$ constitutes all of the dark matter~\cite{Chung:2004nh, Ling:2021zlj, Garcia:2022vwm, Garcia:2023awt}.

Our central finding is that negative field-space curvature systematically enhances gravitational particle production. The physical mechanism proceeds in stages. First, the curved field-space metric modifies the post-inflationary inflaton dynamics: the Christoffel symbol terms source oscillations in the heavy direction $\phi_Q$, while the metric factor $e^{2\phi_Q/L_{\rm fs}}$ amplifies the kinetic energy $G_{IJ}\dot\phi^I\dot\phi^J$ and hence the Ricci scalar $R$. The spectator $\chi$, which couples only to gravity, is affected entirely through the resulting enhancement of the oscillations of $R(\eta)$ and $a(\eta)$ in its mode equation. The net result is an enhancement of the CGPP number density by up to an order of magnitude relative to the flat Starobinsky limit, together with a substantially richer UV interference structure. For the sidetracked attractor, however, this enhanced production efficiency competes with a dramatically reduced inflationary energy scale, leading to a nontrivial dependence of the relic abundance on model parameters.

The remainder of this paper is organized as follows. In Sec.~\ref{sec:multifield} we review the covariant multifield inflation formalism, establishing the notation and perturbation equations used throughout. Section~\ref{sec:benchmark_models} introduces the benchmark models: the flat Starobinsky\,+\,quadratic scan (Sec.~\ref{subsec:starobinsky_quadratic}) and the sidetracked inflation scan (Sec.~\ref{subsec:sidetracked}), including the CMB observables and field-space trajectories. Section~\ref{sec:cgpp} presents the CGPP formalism (Sec.~\ref{subsec:cgpp_formalism}), the general features of multifield particle production (Sec.~\ref{cgpp:multifield}), and the numerical spectra for the SQ (Sec.~\ref{subsec:cgpp_SQ}) and sidetracked (Sec.~\ref{subsec:cgpp_ST}) models. The dark matter relic abundance and isocurvature constraints are discussed in Secs.~\ref{subsec:relicabundance} and~\ref{subsec:iso}. We conclude in Sec.~\ref{sec:conclusions}. Throughout, we work in natural units $\hbar = c = 1$ with the reduced Planck mass $M_{\rm Pl} = (8\pi G)^{-1/2} \simeq 2.435 \times 10^{18}\,$GeV and metric signature $(+,-,-,-)$.

\section{Multifield Inflation}
\label{sec:multifield}
In this section, we establish the covariant formalism for inflation driven by $N$ scalar fields $\phi^I$ ($I=1,\dots,N$) propagating on a curved field-space manifold with metric $G_{IJ}(\phi^1,\ldots,\phi^N)$. This framework provides the natural starting point for studying gravitational particle production in UV-motivated scenarios, including supergravity~\cite{Ferrara:2013rsa, Kallosh:2013hoa, Ellis:2013nxa}, string compactifications~\cite{Baumann:2014nda}, and nonminimally coupled theories~\cite{Kaiser:2010ps, Greenwood:2012aj, Kaiser:2012ak, Schutz:2013fua}, where nontrivial kinetic structures and turning trajectories are generic features. We adopt the covariant approaches to background dynamics and perturbations developed in Refs.~\cite{Sasaki:1995aw, Gordon:2000hv, GrootNibbelink:2001qt, Byrnes:2006fr, Lalak:2007vi, Peterson:2010np, Peterson:2010mv, Gong:2011uw, Kaiser:2012ak, Ellis:2014opa} (see also the reviews~\cite{Wands:2007bd, Malik:2008im, Gong:2016qmq}).

\subsection{Action and Background Dynamics}
\label{subsec:action_covariant}
We consider Einstein gravity minimally coupled to $N$ scalar fields,
\begin{equation}
S \; = \; \int {\rm d}^4x\,\sqrt{-g}\,
\left[
\frac{M_{\rm Pl}^2}{2}\,R
+ \frac{1}{2}\,G_{IJ}(\phi^1,\ldots,\phi^N)\,g^{\mu\nu}\partial_\mu\phi^I\partial_\nu\phi^J
- V(\phi^1,\ldots,\phi^N) \right] \, ,
\label{eq:action_multifield}
\end{equation}
where $R$ is the spacetime Ricci scalar, $V(\phi^1,\ldots,\phi^N)$ is the multifield scalar potential, and $G_{IJ}(\phi^1,\ldots,\phi^N)$ is the field-space metric, i.e.\ a symmetric, positive-definite tensor that endows the scalar manifold $\mathcal{M}$ with a Riemannian structure. For a canonical single field, $G_{IJ} = \delta_{IJ}$, and the action reduces to the standard single-field form. We note that in our convention $R > 0$ for de~Sitter spacetime.

Treating the fields $\phi^I$ as coordinates on the $N$-dimensional Riemannian manifold $\mathcal{M}$, we define covariant derivatives using the Levi-Civita connection of $G_{IJ}$:
\begin{equation}
\Gamma^{I}{}_{JK} \; = \; \frac{1}{2}\,G^{IL}\left(\partial_J G_{KL} + \partial_K G_{JL} - \partial_L G_{JK}\right) \,,
\label{eq:christoffel}
\end{equation}
where $\partial_I \equiv \partial/\partial\phi^I$. For a scalar function such as the potential, the first and second covariant derivatives read
\begin{equation}
\nabla_I V \; = \; \partial_I V \equiv V_{,I} \,, \qquad \nabla_I \nabla_J V \equiv V_{;IJ} \; = \; \partial_I \partial_J V - \Gamma^{K}{}_{IJ}\,\partial_K V \,.
\label{eq:cov_deriv_V}
\end{equation}
The field-space Riemann curvature tensor is
\begin{equation}
\mathcal{R}^{I}{}_{JKL}
\; = \; \partial_K \Gamma^{I}{}_{JL} - \partial_L \Gamma^{I}{}_{JK}
+ \Gamma^{I}{}_{KM}\Gamma^{M}{}_{JL}
- \Gamma^{I}{}_{LM}\Gamma^{M}{}_{JK} \,,
\label{eq:riemann_fieldspace}
\end{equation}
with the associated Ricci tensor $\mathcal{R}_{IJ} = \mathcal{R}^{K}{}_{IKJ}$ and Ricci scalar $\mathcal{R} = G^{IJ}\mathcal{R}_{IJ}$. This intrinsic curvature enters the fluctuation equations as a purely geometric coupling that can be as important as, or even dominate over, potential-induced masses~\cite{Gong:2011uw, Kaiser:2012ak, Renaux-Petel:2015mga}. For a two-dimensional field space ($N=2$), which is the case we focus on in this work, the Riemann tensor is entirely determined by a single scalar quantity, the Gaussian curvature, $\mathbb{K} = \mathcal{R}/2$, and takes the form
\begin{equation}
\label{eq:riemann_2d}
\mathcal{R}_{IJKL} \; = \; \mathbb{K}\left(G_{IK}G_{JL} - G_{IL}G_{JK}\right) \, .
\end{equation}

We assume a spatially flat Friedmann-Lema\^itre-Robertson-Walker (FLRW) background,
\begin{equation}
{\rm d}s^2 \; = \; {\rm d}t^2 - a(t)^2 \,\delta_{ij}\,{\rm d}x^i {\rm d}x^j \,,
\qquad
H \equiv \dot{a}/a \,,
\label{eq:flrw_metric}
\end{equation}
where $a(t)$ is the scale factor, $H$ is the Hubble parameter, and an overdot denotes differentiation with respect to cosmic time~$t$. We further assume homogeneous field configurations, $\phi^I = \phi^I(t)$.

Varying the action~\eqref{eq:action_multifield} with respect to $\phi^I$ yields the covariant field equations,
\begin{equation}
\mathcal{D}_t \dot\phi^I + 3H\dot\phi^I + G^{IJ}V_{,J} \; = \; 0 \,,
\label{eq:bg_eom}
\end{equation}
where $\mathcal{D}_t$ denotes the covariant time derivative along the field-space trajectory,
\begin{equation}
\mathcal{D}_t A^I \; \equiv \; \dot{A}^I + \Gamma^{I}{}_{JK}\,\dot\phi^J A^K \,,
\label{eq:covariant_time_deriv}
\end{equation}
for any field-space vector $A^I(t)$. Applying this to the field velocity yields
\begin{equation}
\mathcal{D}_t \dot\phi^I \; = \; \ddot\phi^I + \Gamma^{I}{}_{JK}\,\dot\phi^J\dot\phi^K \,.
\label{eq:Dt_phidot}
\end{equation}
This covariant formulation ensures that Eq.~\eqref{eq:bg_eom} transforms as a proper vector on the field-space manifold $\mathcal{M}$ under field reparameterizations $\phi^I \to \tilde{\phi}^I(\phi)$.

The Friedmann constraint and acceleration equations follow from the Einstein equations:
\begin{align}
3 H^2 M_{\rm Pl}^2&= \frac{1}{2}\,\dot\sigma^2 + V \,,
\label{eq:friedmann1}\\[4pt]
\dot{H} &= -\frac{\dot\sigma^2}{2M_{\rm Pl}^2} \, ,
\label{eq:friedmann2}
\end{align}
where we have introduced the total field-space speed,
\begin{equation}
\dot\sigma \; \equiv \; \sqrt{G_{IJ}\,\dot\phi^I\dot\phi^J} \,,
\label{eq:sigma_dot_def}
\end{equation}
which measures the rate at which the background trajectory traverses the scalar manifold. Indices are lowered with the field-space metric, $\dot\phi_I \equiv G_{IJ}\dot\phi^J$.

Equations~\eqref{eq:friedmann1} and~\eqref{eq:friedmann2} are equivalent to the continuity equation $\dot\rho + 3H(\rho + p) = 0$, with energy density $\rho = \frac{1}{2}\dot\sigma^2 + V$ and pressure $p = \frac{1}{2}\dot\sigma^2 - V$.

The first and second Hubble slow-roll parameters are
\begin{equation}
\varepsilon \; \equiv \; -\frac{\dot{H}}{H^2} \; = \; \frac{\dot\sigma^2}{2H^2M_{\rm Pl}^2} \,,
\qquad
\eta \; \equiv \; \varepsilon + \frac{1}{2\varepsilon}\frac{{\rm d}\varepsilon}{{\rm d}N} \,,
\label{eq:slow_roll_params}
\end{equation}
where $N$ is the number of $e$-folds before the end of inflation, defined by
\begin{equation}
N(t) \; \equiv \; \int_{t}^{t_{\rm end}} H(t')\,{\rm d}t' \,,
\qquad
\frac{{\rm d}N}{{\rm d}t} = -H \,.
\label{eq:efolds}
\end{equation}
Here $t_{\rm end}$ denotes the time at which $\varepsilon(t_{\rm end}) = 1$. The number of $e$-folds at the horizon exit of the CMB pivot scale $k_*$ is denoted by $N_* \equiv N(t_*)$, where $k_* = a(t_*)H(t_*)$. For the models considered in this paper, $N_* \simeq 50$--$60$, depending on the details of reheating~\cite{Liddle:2003as, Martin:2010kz}. Inflation occurs when $\varepsilon < 1$, and the expansion is quasi-de~Sitter when $\varepsilon \ll 1$.

In the slow-roll regime ($\varepsilon, |\eta| \ll 1$), Eqs.~\eqref{eq:bg_eom} and~\eqref{eq:friedmann1} reduce to
\begin{equation}
3 H \dot\phi^I \; \simeq \; -G^{IJ}V_{,J} \,,
\qquad
3 H^2 M_{\rm Pl}^2 \; \simeq \; V \,,
\label{eq:slow_roll_approx}
\end{equation}
where we have dropped the $\mathcal{D}_t\dot\phi^I$ term and the kinetic contribution to the energy density, respectively. We note in passing that the first of these approximations breaks down in rapid-turn or sidetracked regimes, where $\mathcal{D}_t\dot\phi^I$ remains of order $H\dot\phi^I$ even when $\varepsilon$ and $|\eta|$ are small. We return to this point in Sec.~\ref{subsec:sidetracked}.

\subsection{Kinematic Basis and the Adiabatic/Entropic Decomposition}
\label{subsec:kinematic_basis}
The physics of multifield inflation is most transparently analyzed by decomposing fluctuations into components parallel and perpendicular to the background field-space trajectory. This kinematic decomposition, introduced in Refs.~\cite{Gordon:2000hv, GrootNibbelink:2001qt}, provides a natural basis that separates the adiabatic fluctuation, which sources the curvature perturbation, from the entropic (isocurvature) fluctuation, which can source the adiabatic mode on superhorizon scales.

The unit tangent vector to the background trajectory is
\begin{equation}
e^I_\sigma \; \equiv \; \frac{\dot\phi^I}{\dot\sigma} \,,
\qquad
G_{IJ}\,e^I_\sigma e^J_\sigma = 1 \,.
\label{eq:tangent_vector}
\end{equation}
For the two-field case ($N=2$), to which we restrict from now on, there is a unique unit normal vector $e^I_s$ satisfying the orthonormality conditions
\begin{equation}
G_{IJ}\,e^I_\sigma e^J_s = 0 \,,
\qquad
G_{IJ}\,e^I_s e^J_s = 1 \,,
\label{eq:normal_vector}
\end{equation}
with the orientation fixed by
\begin{equation}
e^I_s \; = \; \frac{1}{\sqrt{G}}\,\epsilon^{IJ}\,G_{JK}\,e^K_\sigma \,,
\label{eq:normal_explicit}
\end{equation}
where $G \equiv \det G_{IJ}$ and $\epsilon^{IJ}$ is the antisymmetric Levi-Civita symbol (a tensor density) with $\epsilon^{12} = 1$. The factor $1/\sqrt{G}$ promotes it to a proper tensor on $\mathcal{M}$.

The turning rate of the trajectory is defined as the magnitude of the covariant rate of change of the tangent vector,
\begin{equation}
\omega \; \equiv \; \left|\mathcal{D}_t e^I_\sigma\right| \; = \; \sqrt{G_{IJ}\,(\mathcal{D}_t e^I_\sigma)(\mathcal{D}_t e^J_\sigma)} \,,
\label{eq:turning_rate_def}
\end{equation}
which, by Eq.~\eqref{eq:tangent_vector}, is orthogonal to $e^I_\sigma$ and therefore parallel to $e^I_s$:
\begin{equation}
\mathcal{D}_t e^I_\sigma \; = \; \omega\,e^I_s \,.
\label{eq:transport_tangent}
\end{equation}
The turning rate is a central quantity in multifield dynamics: it controls the linear mixing between adiabatic and entropic fluctuations and thereby captures the effects of bending of the background trajectory~\cite{Gordon:2000hv, GrootNibbelink:2001qt, Peterson:2010np}. When $\omega = 0$, the trajectory does not bend in field space and this linear coupling vanishes.

From Eq.~\eqref{eq:transport_tangent}, the Frenet-Serret equations for the two-field kinematic basis read\footnote{For $N > 2$ fields, the kinematic basis $\{e^I_n\}$ ($n = 1,\dots,N$) obeys higher-dimensional Frenet-Serret equations with $N-1$ generalized turning rates. See Ref.~\cite{Peterson:2010np}.}
\begin{equation}
\mathcal{D}_t e^I_\sigma \; = \; \omega\,e^I_s \,,
\qquad
\mathcal{D}_t e^I_s \; = \; -\omega\,e^I_\sigma \,.
\label{eq:frenet_serret}
\end{equation}
The second relation follows from differentiating the orthonormality constraint $G_{IJ}e^I_\sigma e^J_s = 0$ and using metric compatibility of $\mathcal{D}_t$.

Projecting the background equations of motion~\eqref{eq:bg_eom} along $e^I_\sigma$ and $e^I_s$ yields
\begin{align}
\ddot\sigma + 3H\dot\sigma + V_{,\sigma} &= 0 \,,
\label{eq:eom_sigma}\\[4pt]
\omega\,\dot\sigma + V_{,s} &= 0 \,,
\label{eq:eom_turn}
\end{align}
where the projected potential gradients are\footnote{$V_{,\sigma}$ and $V_{,s}$ are directional derivatives of the potential along the tangent and normal directions, not partial derivatives with respect to coordinates $\sigma$ and $s$.}
\begin{equation}
V_{,\sigma} \; \equiv \; e^I_\sigma V_{,I} \,,
\qquad
V_{,s} \; \equiv \; e^I_s V_{,I} \,.
\label{eq:projected_gradients}
\end{equation}
Equation~\eqref{eq:eom_sigma} is identical in form to the single-field Klein-Gordon equation and governs the evolution of the adiabatic speed $\dot\sigma$. Equation~\eqref{eq:eom_turn} is the turning equation: it shows that a nonzero turning rate $\omega \neq 0$ is sourced by the component of the potential gradient perpendicular to the trajectory, $V_{,s}$. Non-bending trajectories satisfy $\omega = 0$ and therefore $V_{,s}=0$, so the potential gradient is everywhere tangent to the motion.

\subsection{Primordial Power Spectra and Observables}
\label{subsec:observables}

Since the main focus of this work is gravitational particle production rather than the detailed phenomenology of inflationary perturbations, we restrict ourselves to the minimal set of equations needed to compute the scalar and tensor observables.

The scalar quantity of phenomenological interest is the comoving curvature perturbation $\mathcal{R}_\mathbf{k}$ (not to be confused with the field-space Ricci tensor $\mathcal{R}_{IJ}$ introduced in Sec.~\ref{subsec:action_covariant}),
\begin{equation}
\mathcal{R}_{\mathbf{k}} \;=\; \frac{H}{\dot{\sigma}}\,Q_{\sigma,\mathbf{k}} \,,
\label{eq:R_from_Qsigma}
\end{equation}
where $Q_{\sigma,\mathbf{k}} \equiv e_{\sigma I} Q^I_{\mathbf{k}}$ and $Q_{s,\mathbf{k}} \equiv e_{s I} Q^I_{\mathbf{k}}$ are the adiabatic and entropic projections of the gauge-invariant Mukhanov--Sasaki field fluctuation~\cite{Sasaki:1986hm,Mukhanov:1988jd} onto the kinematic basis introduced in Sec.~\ref{subsec:kinematic_basis}. A distinctive feature of multifield inflation is the existence of entropic (isocurvature) perturbations, i.e., field fluctuations orthogonal to the background trajectory, which have no analogue in the single-field case. The dimensionless isocurvature perturbation is defined by~\cite{Gordon:2000hv,GrootNibbelink:2001qt}
\begin{equation}
\mathcal{S}_{\mathbf{k}} \;\equiv\; \frac{H}{\dot\sigma}\,Q_{s,\mathbf{k}} \,,
\label{eq:isocurvature_def}
\end{equation}
in direct analogy with the curvature perturbation~\eqref{eq:R_from_Qsigma}. While $\mathcal{R}_\mathbf{k}$ is conserved on superhorizon scales in the single-field limit, a nonzero turning rate $\omega$ transfers power from $\mathcal{S}_\mathbf{k}$ to $\mathcal{R}_\mathbf{k}$ after horizon crossing, potentially enhancing the curvature power spectrum at superhorizon scales~\cite{Gordon:2000hv,Amendola:2001ni,Wands:2002bn}. Furthermore, any residual isocurvature component that survives until late times is constrained by CMB observations~\cite{Planck:2018jri}. We quantify these constraints for each model in subsequent sections.

For the two-field case ($N=2$) studied in this work, the linearized equations of motion for the adiabatic and entropic mode functions take the form~\cite{Gordon:2000hv,GrootNibbelink:2001qt,Gong:2011uw,Kaiser:2012ak,McDonough:2020gmn}\footnote{For $N>2$ fields, the single entropic mode $Q_s$ is replaced by $N-1$ entropic components $Q_{s_n}$ ($n = 2,\ldots,N$) with additional couplings among them. See, e.g., Refs.~\cite{Peterson:2010np,Peterson:2010mv}.}
\begin{align}
\ddot{Q}_{\sigma,\mathbf{k}} + 3H\dot{Q}_{\sigma,\mathbf{k}}
+ \left(\frac{k^2}{a^2} + \mu_\sigma^2\right) Q_{\sigma,\mathbf{k}}
&=
2\frac{{\rm d}}{{\rm d}t}\!\left(\omega Q_{s,\mathbf{k}}\right)
-2\left(\frac{V_{,\sigma}}{\dot{\sigma}} + \frac{\dot{H}}{H}\right)\omega\, Q_{s,\mathbf{k}} \,,
\label{eq:Qsigma_eq}\\[6pt]
\ddot{Q}_{s,\mathbf{k}} + 3H\dot{Q}_{s,\mathbf{k}}
+ \left(\frac{k^2}{a^2} + \mu_s^2\right) Q_{s,\mathbf{k}}
&=
4 M_{\rm Pl}^2 \frac{\omega}{\dot{\sigma}} \frac{k^2}{a^2} \Psi_{\mathbf{k}} \,,
\label{eq:Qs_eq}
\end{align}
where $\Psi_\mathbf{k}$ is the Bardeen potential~\cite{Bardeen:1980kt}, related to the curvature perturbation and field velocities through the Einstein constraint equations,\footnote{Explicitly, in longitudinal (Newtonian) gauge the scalar metric perturbations take the form ${\rm d}s^2 = (1+2\Psi){\rm d}t^2 - a^2(1-2\Psi)\delta_{ij}{\rm d}x^i{\rm d}x^j$. The equality of the two scalar potentials $\Phi = \Psi$ follows from the absence of anisotropic stress at linear order in scalar perturbations~\cite{Mukhanov:1990me,Malik:2008im}.} so that Eqs.~\eqref{eq:Qsigma_eq}--\eqref{eq:Qs_eq} form a closed system once the Einstein constraints are imposed. The effective mass-squared terms are
\begin{align}
\mu_\sigma^2
&=
V_{;\sigma\sigma}
-\omega^2
-\frac{1}{M_{\rm Pl}^2 a^3}\frac{{\rm d}}{{\rm d}t}
\left(\frac{a^3\dot{\sigma}^2}{H}\right) ,
\label{eq:mu_sigma}\\[4pt]
\mu_s^2
&=
V_{;ss}
+3\omega^2 + \mathcal{R}_{s\sigma s\sigma}\,\dot{\sigma}^2\,.
\label{eq:mu_s}
\end{align}
Here $V_{;\sigma\sigma}\equiv e_\sigma^I e_\sigma^J V_{;IJ}$ and $V_{;ss}\equiv e_s^I e_s^J V_{;IJ}$ are the projected second covariant derivatives of the potential, and $\mathcal{R}_{s\sigma s\sigma}\equiv
e_s^I e_\sigma^J e_s^K e_\sigma^L \mathcal{R}_{IJKL}$ is the field-space sectional curvature in the plane spanned by the tangent and normal directions. For the two-dimensional case, $\mathcal{R}_{s\sigma s\sigma} = \mathbb{K}$. The coupling between the adiabatic and entropic sectors is controlled by the turning rate $\omega$. When $\omega=0$, the two sectors decouple at linear order. Even when $\omega \neq 0$, the explicit source term on the right-hand side of Eq.~\eqref{eq:Qs_eq} is suppressed on superhorizon scales by the factor $k^2/a^2$.

Deep inside the horizon, $k\gg aH$ and $k/a \gg |\mu_\sigma|, |\mu_s|$, the perturbations are initialized in the Bunch--Davies vacuum,
\begin{equation}
Q_{\sigma,\mathbf{k}},\;Q_{s,\mathbf{k}}
\;\to\;
\frac{e^{-ik\tau}}{a\sqrt{2k}} \,,
\label{eq:BD_modes}
\end{equation}
where $\tau$ is conformal time, defined by ${\rm d}\tau = {\rm d}t/a$.
The dimensionless scalar power spectrum is defined by
\begin{equation}
\langle \mathcal R_{\mathbf{k}}\,\mathcal R_{\mathbf{k}'}^* \rangle
\;=\;
(2\pi)^3\delta^{(3)}(\mathbf{k}-\mathbf{k}')
\,\frac{2\pi^2}{k^3}\,\mathcal P_{\mathcal R}(k)\,.
\label{eq:PR_def}
\end{equation}
Once $\mathcal P_{\mathcal R}(k)$ is obtained numerically, the scalar spectral index is
\begin{equation}
n_s(k)-1
\;\equiv\;
\frac{{\rm d}\ln \mathcal P_{\mathcal R}(k)}{{\rm d}\ln k}\,,
\label{eq:ns_def}
\end{equation}
and is evaluated at the pivot scale $k_*=0.05\,{\rm Mpc}^{-1}$ unless stated otherwise. The amplitude is normalized to the measured value~\cite{Planck:2018jri},
\begin{equation}
\mathcal P_{\mathcal R}(k_*) \;=\; A_s \; = \;2.10 \times 10^{-9} \,.
\label{eq:As_normalization}
\end{equation}

Tensor perturbations are described by the two polarization states of the transverse-traceless metric fluctuation, $h_{\mathbf{k},\gamma}$ with $\gamma=+,\times$, satisfying
\begin{equation}
\ddot h_{\mathbf{k},\gamma}
+3H\dot h_{\mathbf{k},\gamma}
+\frac{k^2}{a^2}h_{\mathbf{k},\gamma}
=0 \,.
\label{eq:tensor_eq}
\end{equation}
The corresponding tensor power spectrum is defined by
\begin{equation}
\sum_{\gamma=+,\times}
\langle h_{\mathbf{k},\gamma}\,h_{\mathbf{k}',\gamma}^* \rangle
\;=\;
(2\pi)^3\delta^{(3)}(\mathbf{k}-\mathbf{k}')
\,\frac{2\pi^2}{k^3}\,\mathcal P_T(k)\,,
\label{eq:PT_def}
\end{equation}
and, at leading order in slow roll, is given by
\begin{equation}
\mathcal P_T(k)
\;=\;
\left.\frac{2H^2}{\pi^2 M_{\rm Pl}^2}\right|_{k=aH} \,.
\label{eq:PT_slowroll}
\end{equation}
The tensor-to-scalar ratio is then
\begin{equation}
r(k)
\;\equiv\;
\frac{\mathcal P_T(k)}{\mathcal P_{\mathcal R}(k)} \,,
\label{eq:r_def}
\end{equation}
again typically quoted at the pivot scale $k_*$.
In the multifield models studied here, we determine $\mathcal P_{\mathcal R}(k)$ numerically by evolving Eqs.~\eqref{eq:Qsigma_eq} and \eqref{eq:Qs_eq} on the background solutions described above, and then extract $n_s$ and $r$ directly from the resulting scalar and tensor spectra. This automatically incorporates any sourcing of the curvature perturbation by entropic modes during and after horizon crossing.

The current CMB constraints are consistent with a red-tilted scalar spectrum,
\begin{equation}
n_s = 0.9665 \pm 0.0075 \qquad (95\%~{\rm CL}),
\label{eq:ns_constraint}
\end{equation}
for the \textit{Planck} TT,TE,EE+lowE+lensing+BAO combination~\cite{Planck:2018vyg},
and an upper bound on the tensor-to-scalar ratio,
\begin{equation}
r_{0.05} < 0.036 \qquad (95\%~{\rm CL}),
\label{eq:r_constraint}
\end{equation}
from the BK18 analysis~\cite{BICEP:2021xfz}. The running of the scalar spectral index is constrained to be
\begin{equation}
\frac{{\rm d}n_s}{{\rm d}\ln k} = -0.0045 \pm 0.0067 \,,
\label{eq:running_constraint}
\end{equation}
for \textit{Planck}+BAO~\cite{Planck:2018jri}. We verify the consistency of each model studied in this work with these bounds.

The dark matter field $\chi$ introduced in Sec.~\ref{sec:cgpp} is a spectator that does not participate in the inflationary dynamics. Its role is purely passive: the multifield inflaton sector determines the background evolution $a(t)$, $H(t)$, and the Ricci scalar $R(t)$, which in turn govern the gravitational production of $\chi$ quanta through the time-dependent geometry.

\section{Benchmark Models}
\label{sec:benchmark_models}
In this section, we apply the general formalism developed in Sec.~\ref{sec:multifield} to concrete two-field models that serve as benchmarks for gravitational particle production. Our focus is on genuinely multifield dynamics, in which both scalar fields participate nontrivially in the background evolution and the field-space trajectory can undergo significant turning. These models therefore provide a useful setting in which to study how multifield effects modify the inflationary dynamics and, in turn, the gravitational production of spectator particles.

\subsection{Starobinsky + Quadratic Inflation}
\label{subsec:starobinsky_quadratic}
We begin with a two-field model that combines the Starobinsky plateau~\cite{Starobinsky:1980te} with quadratic inflation~\cite{Linde:1983gd}. In the Einstein frame, the two fields $\phi_S$ (the scalaron arising from the $R + R^2$ modification of gravity) and $\phi_Q$ (a canonical scalar with a quadratic potential) have canonical kinetic terms and no direct couplings, interacting only via their joint contribution to the cosmological background through Eqs.~\eqref{eq:friedmann1}--\eqref{eq:friedmann2}. The field-space metric is therefore flat,
\begin{equation}
G_{IJ} \; = \; \delta_{IJ} \,, \qquad (\phi^1, \phi^2) \equiv (\phi_S, \phi_Q) \,,
\label{eq:flat_metric_SQ}
\end{equation}
so that the Christoffel symbols vanish, $\Gamma^I{}_{JK} = 0$, and the field-space Riemann tensor is identically zero, $\mathcal{R}^I{}_{JKL} = 0$. The scalar potential is~\cite{Ellis:2013nxa}
\begin{equation}
V_{\rm SQ}(\phi_S, \phi_Q) \; = \; \frac{3}{4}\, m_S^2 M_{\rm Pl}^2 \left( 1 - e^{-\sqrt{2/3}\, \phi_S/M_{\rm Pl}} \right)^2 + \frac{1}{2}\, m_Q^2 \phi_Q^2 \,,
\label{eq:V_SQ}
\end{equation}
where $m_S$ sets the scale of the Starobinsky plateau and $m_Q$ is the mass of the $\phi_Q$ field.
The potential gradients are
\begin{equation}
V_{{\rm SQ},\phi_S} \; = \; \sqrt{\frac{3}{2}}\, m_S^2 M_{\rm Pl} \left( 1 - e^{-\sqrt{2/3}\, \phi_S/M_{\rm Pl}} \right) e^{-\sqrt{2/3}\, \phi_S/M_{\rm Pl}}\,, \qquad V_{{\rm SQ},\phi_Q} \; = \; m_Q^2 \phi_Q\,,
\label{eq:V_SQ_grad}
\end{equation}
and the Hessian components are
\begin{equation}
\begin{aligned}
V_{{\rm SQ},\phi_S\phi_S} &= m_S^2 \left( 2\, e^{-2\sqrt{2/3}\, \phi_S/M_{\rm Pl}} - e^{-\sqrt{2/3}\, \phi_S/M_{\rm Pl}} \right),\\
V_{{\rm SQ},\phi_Q\phi_Q} &= m_Q^2\,, \qquad V_{{\rm SQ},\phi_S \phi_Q} \; = \; 0 \,.
\end{aligned}
\label{eq:V_SQ_hessian}
\end{equation}
The vanishing of the mixed derivative $V_{{\rm SQ},\phi_S \phi_Q}$ reflects the separable (sum of potentials) structure of Eq.~\eqref{eq:V_SQ}. This does not preclude multifield dynamics: when both fields are displaced from the minimum, the gradient $\nabla V$ generally fails to align with the trajectory $\dot\phi^I$, so $V_{,s} \neq 0$ and the trajectory bends in field space.

Since the field-space metric is flat, the covariant background equations~\eqref{eq:bg_eom} reduce to two decoupled Klein-Gordon equations,
\begin{equation}
\ddot\phi_S + 3H\dot\phi_S + V_{{\rm SQ},\phi_S} = 0 \,, \qquad
\ddot{\phi}_Q + 3H\dot{\phi}_Q + V_{{\rm SQ},\phi_Q} = 0 \,,
\label{eq:bg_SQ}
\end{equation}
coupled to each other only through the Hubble parameter $H$ via the Friedmann equations~\eqref{eq:friedmann1}--\eqref{eq:friedmann2}, with the field-space speed given by $\dot\sigma^2 = \dot\phi_S^2 + \dot\phi_Q^2$.

For the flat field-space metric~\eqref{eq:flat_metric_SQ}, the kinematic basis vectors defined in Eqs.~\eqref{eq:tangent_vector}--\eqref{eq:normal_explicit} and the turning rate~\eqref{eq:turning_rate_def} simplify to
\begin{equation}
e_\sigma^I \; = \; \frac{1}{\dot\sigma}(\dot\phi_S, \dot\phi_Q) \,, \qquad
e_s^I \; = \; \frac{1}{\dot\sigma}(\dot\phi_Q, -\dot\phi_S)\,, \qquad
\omega \; = \; \frac{|\dot\phi_S \ddot\phi_Q - \dot\phi_Q \ddot\phi_S|}{\dot\sigma^2} \,,
\label{eq:basis_SQ}
\end{equation}
where the orientation of $e_s^I$ follows from Eq.~\eqref{eq:normal_explicit} with $\epsilon^{12} = +1$. Equivalently, the turning rate can be expressed in terms of the potential gradients using the turning equation~\eqref{eq:eom_turn},
\begin{equation}
\omega \; = \; \frac{|V_{{\rm SQ},\phi_Q}\, \dot\phi_S - V_{{\rm SQ},\phi_S}\, \dot\phi_Q|}{\dot\sigma^2} \; = \; \frac{|m_Q^2 \phi_Q\, \dot\phi_S - V_{{\rm SQ},\phi_S}\, \dot\phi_Q|}{\dot\sigma^2} \,.
\label{eq:omega_SQ}
\end{equation}
Significant turning ($\omega/H \gtrsim 1$) occurs when the transverse potential gradient $V_{,s} = e_s^I V_{,I}$ is comparable to the Hubble friction term $\sim H\dot\sigma$ in Eq.~\eqref{eq:eom_sigma}.

The number of $e$-folds from horizon exit to the end of inflation is, in the slow-roll approximation,
\begin{equation}
N_* \simeq \frac{3}{4}\left[e^{\sqrt{2/3}\,\phi_{S*}/M_{\rm Pl}}
- e^{\sqrt{2/3}\,\phi_{S,\rm end}/M_{\rm Pl}}
- \sqrt{\frac{2}{3}}\frac{\phi_{S*}-\phi_{S,\rm end}}{M_{\rm Pl}}\right]
\;\simeq\; \frac{3}{4}\,e^{\sqrt{2/3}\,\phi_{S*}/M_{\rm Pl}} \,,
\label{eq:Nstar_starobinsky}
\end{equation}
where the final approximation holds for $\phi_{S*}\gg M_{\rm Pl}$. Inverting and substituting into the slow-roll parameters yields the well-known predictions~\cite{Starobinsky:1980te, Kallosh:2013hoa, Ellis:2013nxa}
\begin{equation}
n_s \; \simeq \; 1 - \frac{2}{N_*} \,, \qquad r \; \simeq \; \frac{12}{N_*^2} \,.
\label{eq:starobinsky_ns_r}
\end{equation}
For $N_* = 55$, these give $n_s \simeq 0.964$ and $r \simeq 0.004$, in excellent agreement with CMB constraints~\cite{Planck:2018jri, BICEP:2021xfz}.\footnote{For a recent discussion of Starobinsky inflation in light of the ACT and SPT data, see Ref.~\cite{Ellis:2025zrf}.}

The overall mass scale $m_S$ is fixed by the amplitude normalization~\eqref{eq:As_normalization}. In the single-field slow-roll limit, the scalar power spectrum~\eqref{eq:PR_def} is given by
\begin{equation}
\mathcal{P}_{\mathcal{R}}(k_*) \; = \; \left.\frac{H^2}{8\pi^2\varepsilon\,M_{\rm Pl}^2}\right|_* \, .
\label{eq:PR_slowroll_SQ}
\end{equation}
Using the Starobinsky potential~\eqref{eq:V_SQ} with $\phi_Q = 0$, the normalization condition $\mathcal{P}_\mathcal{R}(k_*) = A_s$ yields
\begin{equation}
m_S \; \simeq \; \sqrt{\frac{24\pi^2 A_s}{N_*^2}}\;M_{\rm Pl} \;\simeq\; 1.3 \times 10^{-5}\,M_{\rm Pl} \;\simeq\; 3.1 \times 10^{13}\,\mathrm{GeV} \,,
\label{eq:m1_starobinsky}
\end{equation}
for $N_* = 55$. In the numerical analysis below, we determine $m_S$ more precisely by solving the full background equations and imposing the exact normalization at the pivot scale.

When the second field is displaced from the origin, $\phi_Q(t_0) \neq 0$, two-field dynamics emerges. The field-space trajectory curves as both fields evolve toward the global minimum at $(\phi_S, \phi_Q) = (0, 0)$. The character of the evolution depends on the ratio of $m_Q$ to the inflationary Hubble scale $H_{*} \equiv H(t_*)$, where $t_*$ is the time at which the CMB pivot scale exits the horizon, $N_* = 55$ $e$-folds before the end of inflation:
\begin{itemize}
\item \emph{Heavy field} ($m_Q \gg H_{*}$): The field $\phi_Q$ rapidly relaxes toward $\phi_Q = 0$, and the trajectory converges to the single-field Starobinsky attractor. During the relaxation phase, the slow-roll parameter $\varepsilon$ exhibits transient deviations from its single-field value, but these decay within a few $e$-folds. In the extreme heavy limit, the heavy field $\phi_Q$ can be integrated out, yielding an effective single-field theory~\cite{Achucarro:2012sm}.

\item \emph{Light or intermediate field} ($m_Q \lesssim H_{*}$): The field $\phi_Q$ remains dynamical during a significant portion of inflation. Both fields contribute to the inflationary trajectory, which curves in the $(\phi_S, \phi_Q)$ plane. This regime features nontrivial turning ($\omega/H \neq 0$), adiabatic-entropic mode mixing through the coupled perturbation equations~\eqref{eq:Qsigma_eq}--\eqref{eq:Qs_eq}, and potentially observable consequences such as scale-dependent features, residual isocurvature perturbations~\cite{Planck:2018jri}, or enhanced non-Gaussianity~\cite{Byrnes:2008wi, Peterson:2010mv}.
\end{itemize}

\begin{figure}[t]
\centering
\includegraphics[width=.85\textwidth]{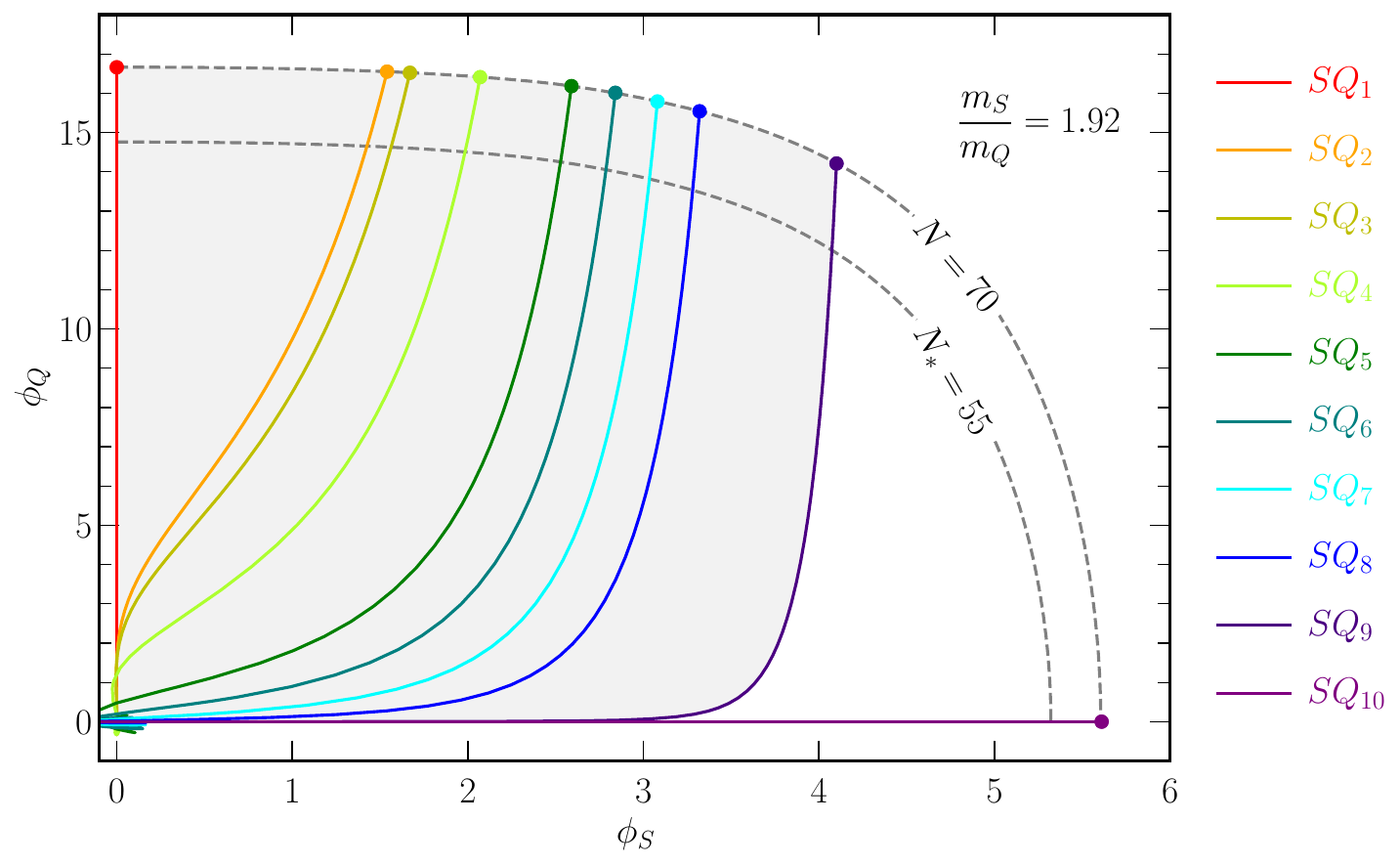}
\caption{Field-space trajectories for the Starobinsky$+$quadratic model Eq.~\eqref{eq:V_SQ} with mass ratio $m_S/m_Q = 1.92$. The ten benchmark models SQ$_1$--SQ$_{10}$ interpolate between pure quadratic inflation (SQ$_1$, red, along $\phi_S=0$) and pure Starobinsky inflation (SQ$_{10}$, purple, along $\phi_Q=0$). Filled circles mark the initial field values at $N_{\rm tot} = 70$ $e$-folds before the end of inflation. The dashed curves indicate the contours of constant total $e$-folds $N_{\rm tot} = 70$ (outer) and $N_* = 55$ (inner). The light gray shaded region is excluded by the joint Planck+BK18 constraints on $n_s$ and $r$. All fields are in units of $M_{\rm Pl}$.}
\label{fig:SQ_trajectories}
\end{figure}
For multifield evolution with significant turning ($\omega/H \gtrsim 1$), the simple slow-roll relations~\eqref{eq:starobinsky_ns_r} are no longer reliable because superhorizon evolution sourced by entropic perturbations can modify $\mathcal{R}_\mathbf{k}$ after horizon exit~\cite{Gordon:2000hv, Wands:2002bn}. In such cases, we compute the full power spectrum by solving the coupled perturbation equations~\eqref{eq:Qsigma_eq}-\eqref{eq:Qs_eq} from subhorizon initialization~\eqref{eq:BD_modes} through to late times when the modes have frozen out, and extract the observables $(n_s, r)$ directly from the numerical spectra via Eqs.~\eqref{eq:ns_def} and \eqref{eq:r_def}.

To map out the interpolation between the pure quadratic and pure Starobinsky limits, we fix the mass ratio $m_S/m_Q = 1.92$, the value for which both single-field limits independently reproduce the observed scalar amplitude $A_s = 2.10 \times 10^{-9}$ at $N_* = 55$. This is close to the slow-roll prediction $m_S/m_Q = \sqrt{4} = 2$, obtained from Eq.~\eqref{eq:m1_starobinsky} and the analogous quadratic result $m_Q \simeq M_{\rm Pl}\sqrt{6\pi^2 A_s/N_*^2}$. The few percent reduction follows from solving the full background equations beyond the leading slow-roll approximation. We select ten benchmark initial conditions $(\phi_{S,i}, \phi_{Q,i})$ lying on the $N_{\rm tot} = 70$ $e$-fold contour, i.e., the locus of initial field values that yield $70$ $e$-folds of inflation.\footnote{The total number of $e$-folds $N_{\rm tot}$ exceeds $N_*$ to ensure that the pivot scale $k_*$ exits the horizon well after any initial transients have decayed. CMB observables are evaluated at $N_* = 55$ $e$-folds before the end of inflation.} The resulting field-space trajectories are shown in Fig.~\ref{fig:SQ_trajectories}. Model SQ$_1$ (red, leftmost) corresponds to pure quadratic inflation with $\phi_{S,i} = 0$, where the trajectory evolves entirely along the $\phi_Q$ direction. Model SQ$_{10}$ (purple, bottom) corresponds to pure Starobinsky inflation with $\phi_{Q,i} = 0$, evolving along the $\phi_S$ axis. Intermediate models SQ$_2$-SQ$_9$ interpolate smoothly between these limits, with progressively more Starobinsky-like character as $\phi_{S,i}$ increases. The filled circles mark the initial field values at $N_{\rm tot} = 70$, and the dashed curves indicate the $N_{\rm tot} = 70$ and $N_* = 55$ contours.

For each benchmark, we evolve the background equations~\eqref{eq:bg_SQ} together with the Friedmann equations~\eqref{eq:friedmann1}-\eqref{eq:friedmann2} forward in time until $\varepsilon(t_{\rm end}) = 1$. Initial velocities are set using the slow-roll attractor, $3H\dot\phi_{I,i} \simeq -V_{,I}|_i$, to minimize transient effects. The Starobinsky-sector mass $m_S$ is adjusted in each model to satisfy the amplitude normalization $\mathcal{P}_\mathcal{R}(k_*) = A_s$~\eqref{eq:As_normalization}, and the CMB observables $n_s$ and $r$ are evaluated at $N_* = 55$ $e$-folds before the end of inflation. The results are summarized in Table~\ref{tab:SQ_runs}.

\begin{table}[t]
\centering
\renewcommand{\arraystretch}{1.3}
\begin{tabular}{l c c c c c c}
\hline\hline
Model & $\phi_{S,i}/M_{\rm Pl}$ & $\phi_{Q,i}/M_{\rm Pl}$ & $m_S/M_{\rm Pl}$ & $n_s$ & $r$ \\
\hline
\ding{55}\, SQ$_{1}$ (quadratic)       & $0$       & $16.66$  & $1.229 \times 10^{-5}$ & $0.9645$ & $0.144$\phantom{0} \\
\ding{55}\, SQ$_{2}$                   & $1.54$    & $16.55$  & $1.222 \times 10^{-5}$ & $0.9640$ & $0.141$\phantom{0} \\
\ding{55}\, SQ$_{3}$                   & $1.67$    & $16.52$  & $1.218 \times 10^{-5}$ & $0.9641$ & $0.140$\phantom{0} \\
\ding{55}\, SQ$_{4}$                   & $2.07$    & $16.41$  & $1.194 \times 10^{-5}$ & $0.9650$ & $0.133$\phantom{0} \\
\ding{55}\, SQ$_{5}$                   & $2.59$    & $16.18$  & $1.100 \times 10^{-5}$ & $0.9675$ & $0.109$\phantom{0} \\
\ding{55}\, SQ$_{6}$                   & $2.84$    & $16.01$  & $1.046 \times 10^{-5}$ & $0.9678$ & $0.096$\phantom{0} \\
\ding{55}\, SQ$_{7}$                   & $3.08$    & $15.79$  & $9.709 \times 10^{-6}$ & $0.9668$ & $0.080$\phantom{0} \\
\ding{55}\, SQ$_{8}$                   & $3.32$    & $15.54$  & $8.898 \times 10^{-6}$ & $0.9651$ & $0.065$\phantom{0} \\
$\checkmark$\,SQ$_{9}$                 & $4.10$    & $14.21$  & $6.205 \times 10^{-6}$ & $0.9695$ & $0.025$\phantom{0} \\
$\checkmark$\,SQ$_{10}$ (Starobinsky)  & $5.61$    & $0$      & $1.226 \times 10^{-5}$ & $0.9659$ & $0.0035$ \\
\hline\hline
\end{tabular}
\caption{
Starobinsky$+$quadratic model with mass ratio $m_S/m_Q = 1.92$ held fixed across all benchmarks. Initial conditions $(\phi_{S,i}, \phi_{Q,i})$ are chosen on the $N_{\rm tot} = 70$ contour, and CMB observables are evaluated at $N_* = 55$. In each model, the Starobinsky-sector mass $m_S$ is fixed by the amplitude normalization $\mathcal{P}_\mathcal{R}(k_*) = A_s = 2.10 \times 10^{-9}$, and $m_Q = m_S/1.92$ follows automatically. A check ($\checkmark$) or cross (\ding{55}) next to the model label indicates consistency with the joint Planck\,+\,BK18 constraints~\cite{Planck:2018jri, BICEP:2021xfz}, taking $n_s \in [0.959, 0.974]$ and $r < 0.036$ at 95\% CL.
}
\label{tab:SQ_runs}
\end{table}

The results reveal a clear trend: as the initial conditions shift from the quadratic-dominated regime (SQ$_1$) toward the Starobinsky-dominated regime (SQ$_{10}$), the tensor-to-scalar ratio decreases monotonically from $r \simeq 0.14$ to $r \simeq 0.004$, spanning nearly two orders of magnitude. The scalar spectral index remains in the range $n_s \simeq 0.964$-$0.970$ across all benchmarks, consistent with the Planck constraint~\eqref{eq:ns_constraint}. However, the tensor-to-scalar ratio exceeds the BK18 bound $r < 0.036$~\eqref{eq:r_constraint} for all runs except SQ$_9$ and SQ$_{10}$. This demonstrates that, for this mass ratio, observational viability requires the Starobinsky sector to dominate the inflationary dynamics, restricting the allowed initial displacement of $\phi_Q$ to a narrow window. These benchmarks will serve as explicit test cases for the gravitational production of the dark matter spectator field $\chi$ in Sec.~\ref{sec:cgpp}.

\subsection{Sidetracked Inflation}
\label{subsec:sidetracked}
As a second benchmark, we consider \emph{sidetracked inflation}: an inflationary attractor with sustained turning induced by negative field-space curvature. This mechanism, introduced in Refs.~\cite{Renaux-Petel:2015mga, Garcia-Saenz:2018ifx, Garcia-Saenz:2018vqf} and placed in the broader context of rapid-turn attractors in Refs.~\cite{Brown:2017osf, Bjorkmo:2019qno, Aragam:2021scu}, arises when the geometric contribution to the entropic mass destabilizes the valley floor and drives the system onto a new attractor with nonzero turning. Gravitational particle production in rapid-turn multifield inflation was also studied in Ref.~\cite{Kolb:2022eyn} for two-field models motivated in part by the de~Sitter swampland conjecture, one of which employs rotationally symmetric potentials with field-space angular-momentum conservation. Here, by contrast, we consider the sidetracked attractor generated by geometrical destabilization of a valley, using the same Starobinsky\,+\,quadratic potential as in Sec.~\ref{subsec:starobinsky_quadratic}. This provides a direct comparison with the flat field-space limit: the potential and field content are unchanged, and the field-space curvature $L_{\rm fs}$ is the only new parameter in the benchmark setup.

We use the same field content $\phi^I=(\phi_S,\phi_Q)$ and the same scalar potential~\eqref{eq:V_SQ} as in Sec.~\ref{subsec:starobinsky_quadratic}, but now promote the field-space metric from flat to a hyperbolic geometry of constant negative curvature,
\begin{equation}
G_{IJ}\,{\rm d}\phi^I\,{\rm d}\phi^J
\;=\;
e^{2\phi_Q/L_{\rm fs}}\,{\rm d}\phi_S^2 + {\rm d}\phi_Q^2 \,,
\label{eq:metric_sidetracked}
\end{equation}
where $L_{\rm fs}>0$ is the field-space curvature. 

This is a parametrization of the Poincaré half-plane model of $\mathbb{H}^2$,\footnote{Indeed, under the coordinate transformation $x=\phi_S/L_{\rm fs}$ and $y=e^{-\phi_Q/L_{\rm fs}}$, the field-space line element becomes $
ds_{\rm fs}^2=L_{\rm fs}^2\frac{dx^2+dy^2}{y^2}.
$
Hyperbolic scalar manifolds of this type arise naturally as K\"ahler manifolds in no-scale supergravity and multifield $\alpha$-attractor constructions; see, e.g., Refs.\cite{Ellis:2013xoa,Carrasco:2015rva,Ellis:2019bmm, Baur:2023naq}. We do not attempt here to embed the particular separable potential of Eq.\eqref{eq:V_SQ} into a complete supergravity model, but instead retain the characteristic hyperbolic kinetic geometry as a controlled effective benchmark.} with constant field-space Ricci scalar
\begin{equation}
\mathcal{R}_{\rm fs} \;=\; -\frac{2}{L_{\rm fs}^2} \,.
\label{eq:Rfs_sidetracked}
\end{equation}
The limit $L_{\rm fs}\to\infty$ recovers $\mathcal{R}_{\rm fs}\to 0$ and reduces the model to the flat Starobinsky\,+\,quadratic benchmark of Sec.~\ref{subsec:starobinsky_quadratic}.

The hyperbolic field-space geometry~\eqref{eq:metric_sidetracked} arises naturally in supergravity and string compactifications. In the $\alpha$-attractor framework~\cite{Kallosh:2013hoa, Ferrara:2013rsa, Kallosh:2013yoa, Galante:2014ifa, Ellis:2013nxa, Ellis:2019bmm}, the K\"ahler manifold of the inflaton sector is $\mathrm{SL}(2,\mathbb{R})/\mathrm{U}(1)\simeq \mathbb H^2$, with field-space curvature set by the parameter $\alpha$ according to\footnote{More precisely, for $\mathcal N=1$ supergravity with a single chiral superfield $\Phi$, the K\"ahler potential $K=-3\alpha\,M_{\rm Pl}^2\,\ln\!\left[(\Phi+\bar\Phi)/(2M_{\rm Pl})\right]$ yields a Poincar\'e disk metric with curvature $\mathcal{R}_{\rm fs}=-2/(3\alpha M_{\rm Pl}^2)$~\cite{Ferrara:2013rsa, Kallosh:2015zsa}. The half-plane parametrization~\eqref{eq:metric_sidetracked} is obtained by a field redefinition.}
\begin{equation}
\mathcal{R}_{\rm fs} \;=\; -\frac{2}{3\alpha\,M_{\rm Pl}^2} \,,
\qquad\text{i.e.,}\qquad
L_{\rm fs} \;=\; \sqrt{3\alpha}\,M_{\rm Pl} \,.
\label{eq:Lfs_alpha}
\end{equation}
Representative values are $\alpha=1$, which gives $L_{\rm fs}=\sqrt{3}\,M_{\rm Pl}\simeq 1.73\,M_{\rm Pl}$, $\alpha=1/3$, which gives $L_{\rm fs}=M_{\rm Pl}$, and $\alpha=1/9$, which gives $L_{\rm fs}=M_{\rm Pl}/\sqrt{3}\simeq 0.58\,M_{\rm Pl}$~\cite{Ferrara:2013rsa, Kallosh:2013yoa}. These curvature radii are all $\mathcal O(M_{\rm Pl})$. As we show below, for the mass ratio $m_S/m_Q=1.92$ of Sec.~\ref{subsec:starobinsky_quadratic}, the critical curvature for destabilization is $L_{\rm crit}\simeq 0.02\,M_{\rm Pl}$ [Eq.~\eqref{eq:Lcrit}], corresponding to $\alpha_{\rm crit}\simeq 1.5\times 10^{-4}$. Standard SUGRA constructions with $\alpha=\mathcal O(1)$ are therefore safely in the stable regime for this mass ratio. Reaching the sidetracked regime with $\alpha=\mathcal O(1)$ requires a lighter second field. For example, for $m_S/m_Q=10$ one finds $L_{\rm crit}\simeq 0.11\,M_{\rm Pl}$, corresponding to $\alpha_{\rm crit}\simeq 4\times 10^{-3}$, a range that can arise in string-motivated compactifications with many moduli~\cite{Cicoli:2018kdo}. We therefore consider benchmarks at both mass ratios below.

The nonvanishing Christoffel symbols for the metric~\eqref{eq:metric_sidetracked} are
\begin{equation}
\Gamma^{\phi_S}{}_{\phi_S\phi_Q}
\;=\;
\Gamma^{\phi_S}{}_{\phi_Q\phi_S}
\;=\;
\frac{1}{L_{\rm fs}} \,,
\qquad
\Gamma^{\phi_Q}{}_{\phi_S\phi_S}
\;=\;
-\frac{e^{2\phi_Q/L_{\rm fs}}}{L_{\rm fs}} \,,
\label{eq:christoffel_sidetracked}
\end{equation}
and the independent components of the field-space Riemann tensor are
\begin{equation}
\mathcal{R}^{\phi_S}{}_{\phi_Q\phi_S\phi_Q}
\;=\;
-\frac{1}{L_{\rm fs}^2} \,,
\qquad
\mathcal{R}^{\phi_Q}{}_{\phi_S\phi_Q\phi_S}
\;=\;
-\frac{e^{2\phi_Q/L_{\rm fs}}}{L_{\rm fs}^2} \,.
\label{eq:riemann_sidetracked}
\end{equation}
The fully covariant Riemann tensor takes the form~\eqref{eq:riemann_2d} with Gaussian curvature $\mathbb{K}=\mathcal{R}_{\rm fs}/2=-1/L_{\rm fs}^2$, as required for a space of constant curvature in two dimensions.
The covariant background equations~\eqref{eq:bg_eom} for the metric~\eqref{eq:metric_sidetracked} become
\begin{align}
\ddot\phi_S + 3H\dot\phi_S + \frac{2}{L_{\rm fs}}\,\dot\phi_S\dot\phi_Q
+ e^{-2\phi_Q/L_{\rm fs}}\,V_{{\rm SQ},\phi_S}
&= 0 \,,
\label{eq:bg1_sidetracked}\\[4pt]
\ddot\phi_Q + 3H\dot\phi_Q - \frac{e^{2\phi_Q/L_{\rm fs}}}{L_{\rm fs}}\,\dot\phi_S^2
+ V_{{\rm SQ},\phi_Q}
&= 0 \,,
\label{eq:bg2_sidetracked}
\end{align}
where $V_{{\rm SQ},\phi_S}$ and $V_{{\rm SQ},\phi_Q}$ are the ordinary partial derivatives given in Eq.~\eqref{eq:V_SQ_grad}, and the field-space speed is
\begin{equation}
\dot\sigma^2 = e^{2\phi_Q/L_{\rm fs}}\dot\phi_S^2 + \dot\phi_Q^2 \,.
\label{eq:sigmadot_sidetracked}
\end{equation}
The velocity-dependent term $-(e^{2\phi_Q/L_{\rm fs}}/L_{\rm fs})\dot\phi_S^2$ in the $\phi_Q$ equation acts as an effective centrifugal force induced by the curved geometry: it can push $\phi_Q$ away from the valley floor $\phi_Q=0$ even when the potential gradient $V_{{\rm SQ},\phi_Q}=m_Q^2\phi_Q$ pulls it back, with the destabilizing term growing exponentially as $\phi_Q$ increases.

The stability of the valley floor $\phi_Q=0$ is controlled by the entropic effective mass. Using Eq.~\eqref{eq:mu_s} and evaluating on the valley, where $\phi_Q=0$ and $\omega=0$, together with the sectional curvature
\begin{equation}
\mathcal{R}_{s\sigma s\sigma}
\;=\;
\frac{\mathcal{R}_{\rm fs}}{2}
\;=\;
-\frac{1}{L_{\rm fs}^2}
\label{eq:Rsection_sidetracked}
\end{equation}
from Eq.~\eqref{eq:riemann_2d}, and the background relation $\dot\sigma^2=2\varepsilon H^2 M_{\rm Pl}^2$, we obtain
\begin{equation}
\mu_s^2\big|_{\rm valley}
\;\simeq\;
m_Q^2-\frac{2\varepsilon H^2 M_{\rm Pl}^2}{L_{\rm fs}^2}\,.
\label{eq:mu_s_valley}
\end{equation}
The geometric contribution is negative for $\mathcal{R}_{\rm fs}<0$ and therefore acts against the stabilizing potential mass $m_Q^2$. The valley becomes tachyonically unstable when
\begin{equation}
\frac{2\varepsilon H^2 M_{\rm Pl}^2}{L_{\rm fs}^2}
\;>\;
m_Q^2\,,
\label{eq:destabilization_condition}
\end{equation}
which defines the onset of \emph{geometrical destabilization}~\cite{Renaux-Petel:2015mga, Renaux-Petel:2017dia}. Once this condition is satisfied, the trajectory is driven away from the valley floor and, after a transient, settles onto the \emph{sidetracked attractor}: a new inflationary solution with $\phi_Q\neq 0$ and sustained turning. The attractor displacement of $\phi_Q$ is set by quasi-static balance in Eq.~\eqref{eq:bg2_sidetracked}: the centrifugal contribution $\sim e^{2\phi_Q/L_{\rm fs}}\dot\phi_S^2/L_{\rm fs}$ from the curved kinetic term, which pushes $\phi_Q$ away from the valley, is balanced by the potential restoring force $m_Q^2\phi_Q$, giving $m_Q^2\phi_Q \sim \dot\sigma^2/L_{\rm fs}$. Combined with the turning equation~\eqref{eq:eom_turn}, $\omega\dot\sigma = -V_{,s} \sim -m_Q^2 \phi_Q$, this yields
\begin{equation}
\frac{\omega}{H}\;\sim\;\frac{\sqrt{2\varepsilon}\,M_{\rm Pl}}{L_{\rm fs}}\;=\;\frac{m_Q}{H}\,\frac{L_{\rm crit}}{L_{\rm fs}}\,,
\label{eq:omega_over_H_sidetracked}
\end{equation}
where the second equality uses Eq.~\eqref{eq:Lcrit}. The turning rate becomes of order $H$ (the rapid-turn regime~\cite{Garcia-Saenz:2018ifx, Bjorkmo:2019qno}) once $L_{\rm fs}$ is reduced sufficiently below $L_{\rm crit}$, with the exact threshold depending on the mass ratio $m_Q/H$.

The destabilization condition~\eqref{eq:destabilization_condition} can be expressed as a bound on the curvature:
\begin{equation}
L_{\rm fs}
\;<\;
L_{\rm crit}
\;\equiv\;
\frac{\sqrt{2\varepsilon}\,H\,M_{\rm Pl}}{m_Q} \,,
\label{eq:Lcrit}
\end{equation}
where $\varepsilon$ and $H$ are evaluated at the time of destabilization (in practice, close to the start of the inflationary run for our benchmarks). For $L_{\rm fs}\gg L_{\rm crit}$, the geometry is nearly flat and the dynamics reduces to the Starobinsky\,+\,quadratic case of Sec.~\ref{subsec:starobinsky_quadratic}. For $L_{\rm fs}\lesssim L_{\rm crit}$, the sidetracked attractor is reached and genuinely multifield effects become important.

Because the turning rate is of order $H$ on the sidetracked attractor, the adiabatic and entropic perturbations are strongly coupled through Eqs.~\eqref{eq:Qsigma_eq}--\eqref{eq:Qs_eq}. As in Sec.~\ref{subsec:starobinsky_quadratic}, we compute $\mathcal P_{\mathcal R}(k)$ and $\mathcal P_T(k)$ numerically by evolving the full coupled perturbation system on the background solutions, and extract $n_s$ and $r$ via Eqs.~\eqref{eq:ns_def} and \eqref{eq:r_def}. Relative to the valley solution, the sidetracked regime typically yields a reduced tensor-to-scalar ratio, because sustained turning alters the standard single-field relation between $\varepsilon$ and $r$~\cite{Garcia-Saenz:2018ifx}.

We organize the sidetracked benchmarks by the dimensionless ratio $L_{\rm fs}/L_{\rm crit}$, which universally parametrizes the distance from the destabilization threshold regardless of the mass ratio. We adopt the values
\begin{equation}
L_{\rm fs}/L_{\rm crit} \;\in\; \{\infty,\; 50,\; 20,\; 10,\; 5,\; 2,\; 1,\; 0.7\} \,,
\label{eq:Lfs_scan}
\end{equation}
spanning from the flat limit through the stable, threshold, and marginally sidetracked regimes. These values are studied at two mass ratios:
\begin{itemize}
\item \emph{Series~A} ($m_S/m_Q = 1.92$, $L_{\rm crit} \simeq 0.021\,M_{\rm Pl}$, $\alpha_{\rm crit} \simeq 1.5 \times 10^{-4}$): The same mass ratio as the flat benchmarks of Sec.~\ref{subsec:starobinsky_quadratic}, enabling a direct comparison. The second field has $m_Q \sim H_{*}$, so destabilization requires strongly curved field spaces with curvature radii well below $M_{\rm Pl}$. The flat limit ST$^{\rm A}_1$ reduces to the pure Starobinsky model SQ$_{10}$ of Table~\ref{tab:SQ_runs}.

\item \emph{Series~B} ($m_S/m_Q = 10$, $L_{\rm crit} \simeq 0.111\,M_{\rm Pl}$, $\alpha_{\rm crit} \simeq 4 \times 10^{-3}$): A lighter second field, $m_Q \simeq 0.2\,H_{*}$. The destabilization threshold is more accessible: the $L_{\rm fs}/L_{\rm crit} = 10$ model corresponds to $L_{\rm fs} \simeq 1.1\,M_{\rm Pl}$ ($\alpha \simeq 0.41$), squarely in the SUGRA-motivated range, while the destabilization threshold $\alpha_{\rm crit} \simeq 4 \times 10^{-3}$ falls within the range of values motivated by string compactifications with many moduli~\cite{Cicoli:2018kdo}.
\end{itemize}

\begin{figure}[t]
\centering
\includegraphics[width=0.85\textwidth]{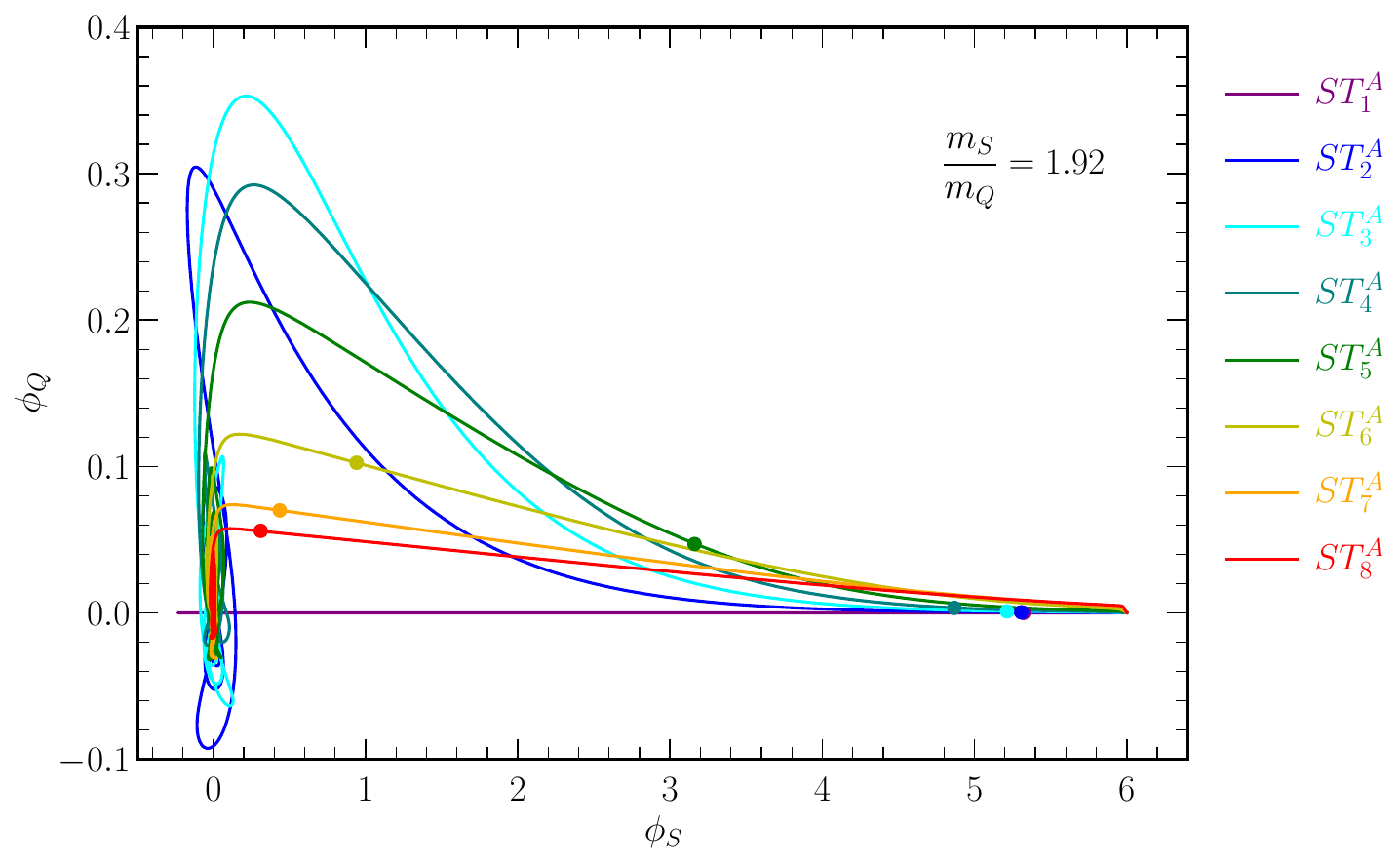}
\caption{Field-space trajectories for the sidetracked inflation benchmarks of Series~A ($m_S/m_Q = 1.92$). ST$^{\rm A}_1$ (purple) is the flat-limit trajectory along $\phi_Q = 0$. As $L_{\rm fs}$ decreases, the trajectory is displaced to progressively larger values of $\phi_Q$ by the curvature-induced centrifugal force. Filled circles mark the field values at the CMB pivot scale. All fields are in units of $M_{\rm Pl}$.}
\label{fig:STA_trajectories}
\end{figure}

\begin{figure}[t]
\centering
\includegraphics[width=0.85\textwidth]{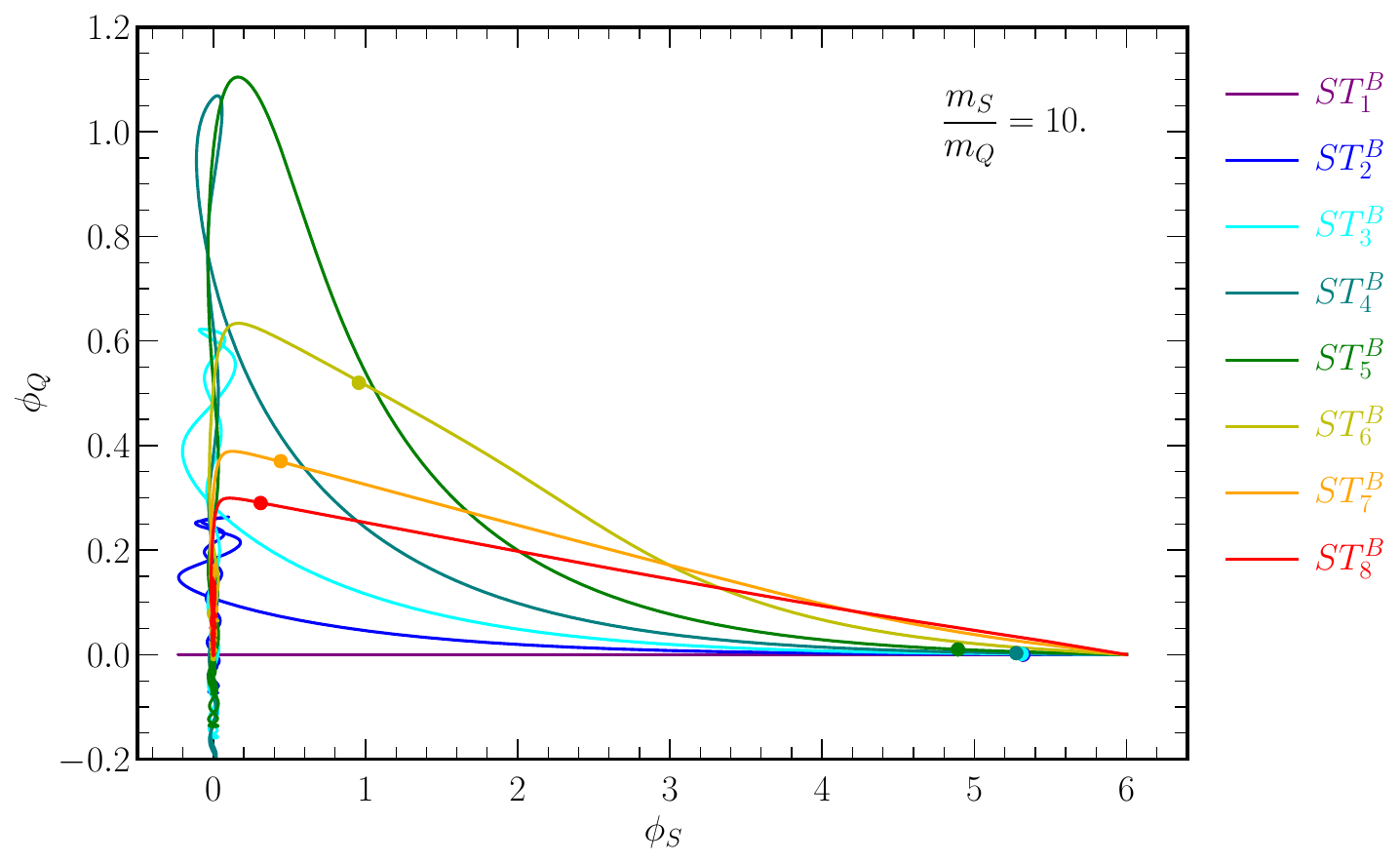}
\caption{Same as Fig.~\ref{fig:STA_trajectories}, but for Series~B ($m_S/m_Q = 10$). Because the $\phi_Q$ field is lighter, the trajectories are more strongly displaced from the valley floor, and the pivot-scale field values occur at larger $\phi_Q$ in the more strongly curved models. }
\label{fig:STB_trajectories}
\end{figure}

All models are initialized at $(\phi_{S,i}, \phi_{Q,i}) = (6, 10^{-4})\,M_{\rm Pl}$, with the small seed displacement $\phi_{Q,i} = 10^{-4}\,M_{\rm Pl}$ chosen to provide a numerical seed for the tachyonic instability in below-threshold runs, and evolved forward until ten Hubble times past the end of inflation, where $\varepsilon(t_{\rm end}) = 1$. For runs in the stable regime ($L_{\rm fs} > L_{\rm crit}$), $\phi_Q$ remains near the valley floor throughout, and the dynamics is a perturbative deformation of single-field Starobinsky inflation. For runs below threshold ($L_{\rm fs} < L_{\rm crit}$), the tachyonic instability amplifies the seed displacement and the trajectory relaxes onto the sidetracked attractor. The CMB observables, evaluated at $N_* = 55$ $e$-folds before the end of inflation, are then determined by the on-attractor dynamics. The Starobinsky-sector mass $m_S$ is adjusted in each model to satisfy the amplitude normalization $\mathcal{P}_\mathcal{R}(k_*) = A_s$, Eq.~\eqref{eq:As_normalization}. The resulting field-space trajectories are shown in Figs.~\ref{fig:STA_trajectories} and~\ref{fig:STB_trajectories}, and the benchmark parameters and CMB observables are summarized in Table~\ref{tab:ST_runs}.

\begin{table}[t]
\centering
\renewcommand{\arraystretch}{1.3}
\setlength{\tabcolsep}{4pt}
\begin{tabular}{l c c c c c c c c}
\hline\hline
Model & $L_{\rm fs}/L_{\rm crit}$ & $L_{\rm fs}/M_{\rm Pl}$ & $\alpha$ & $\phi_{S,*}/M_{\rm Pl}$ & $\phi_{Q,*}/M_{\rm Pl}$ & $m_S/M_{\rm Pl}$ & $n_s$ & $r$ \\
\hline
\multicolumn{9}{c}{\emph{Series A}\; ($m_S/m_Q = 1.92$, \;$L_{\rm crit} \simeq 0.021\,M_{\rm Pl}$)} \\
\hline
$\checkmark\,$ST$^{\rm A}_1$ & $\infty$ & $\infty$ & ---                  & $5.321$ & $0$                   & $1.226 \times 10^{-5}$ & $0.9659$ & $3.5 \times 10^{-3}$ \\
$\checkmark$\,ST$^{\rm A}_2$ & $50$     & $1.069$  & $3.8 \times 10^{-1}$ & $5.307$ & $3.6 \times 10^{-4}$  & $1.237 \times 10^{-5}$ & $0.9690$ & $3.6 \times 10^{-3}$ \\
$\checkmark$\,ST$^{\rm A}_3$ & $20$     & $0.428$  & $6.1 \times 10^{-2}$ & $5.213$ & $1.0\times 10^{-3}$ & $1.296 \times 10^{-5}$ & $0.9680$ & $3.9 \times 10^{-3}$ \\
$\checkmark$\,ST$^{\rm A}_4$ & $10$     & $0.214$  & $1.5 \times 10^{-2}$ & $4.867$ & $3.4 \times 10^{-3}$ & $1.463 \times 10^{-5}$ & $0.9682$ & $5.0 \times 10^{-3}$ \\
$\checkmark$\,ST$^{\rm A}_5$ & $5$      & $0.107$  & $3.8 \times 10^{-3}$ & $3.161$ & $4.7 \times 10^{-2}$  & $1.698 \times 10^{-5}$ & $0.9728$ & $5.9 \times 10^{-3}$ \\
\ding{55}\, ST$^{\rm A}_6$     & $2$      & $0.043$  & $6.1 \times 10^{-4}$ & $0.941$ & $1.0 \times 10^{-1}$  & $1.080 \times 10^{-5}$ & $0.9995$\phantom{0} & $8.1 \times 10^{-4}$ \\
\ding{55}\, ST$^{\rm A}_7$     & $1$      & $0.021$  & $1.5 \times 10^{-4}$ & $0.436$ & $7.0 \times 10^{-2}$  & $1.668 \times 10^{-6}$ & $1.044$\phantom{0} & $6.1 \times 10^{-6}$ \\
\ding{55}\, ST$^{\rm A}_8$ & $0.7$    & $0.015$  & $7.5 \times 10^{-5}$ & $0.311$ & $5.6 \times 10^{-2}$  & $2.994 \times 10^{-7}$ & $1.1032$ & $1.1 \times 10^{-7}$ \\
\hline
\multicolumn{9}{c}{\emph{Series B}\; ($m_S/m_Q = 10$, \;$L_{\rm crit} \simeq 0.111\,M_{\rm Pl}$)} \\
\hline
$\checkmark$\, ST$^{\rm B}_1$ & $\infty$ & $\infty$ & ---                  & $5.321$ & $0$                   & $1.227 \times 10^{-5}$ & $0.9659$ & $3.5 \times 10^{-3}$ \\
$\checkmark$\, ST$^{\rm B}_2$ & $50$     & $5.567$  & $1.0 \times 10^{+1}$ & $5.320$ & $5.8 \times 10^{-4}$  & $1.182 \times 10^{-5}$ & $0.9620$ & $3.3 \times 10^{-3}$ \\
$\checkmark$\, ST$^{\rm B}_3$ & $20$     & $2.227$  & $1.7 \times 10^{+0}$ & $5.311$ & $1.4 \times 10^{-3}$  & $1.003 \times 10^{-5}$ & $0.9660$ & $2.4 \times 10^{-3}$ \\
$\checkmark$\, ST$^{\rm B}_4$ & $10$     & $1.113$  & $4.1 \times 10^{-1}$ & $5.275$ & $2.9 \times 10^{-3}$  & $7.100 \times 10^{-6}$ & $0.9703$ & $1.2 \times 10^{-3}$ \\
$\checkmark$\, ST$^{\rm B}_5$ & $5$      & $0.557$  & $1.0 \times 10^{-1}$ & $4.892$ & $1.0 \times 10^{-2}$  & $4.562 \times 10^{-6}$ & $0.9710$ & $4.8 \times 10^{-4}$ \\
\ding{55}\, ST$^{\rm B}_6$   & $2$      & $0.223$  & $1.7 \times 10^{-2}$ & $0.956$ & $5.2 \times 10^{-1}$  & $2.327 \times 10^{-5}$ & $0.9549$ & $3.8 \times 10^{-3}$ \\
\ding{55}\, ST$^{\rm B}_7$   & $1$      & $0.111$  & $4.1 \times 10^{-3}$ & $0.443$ & $3.7 \times 10^{-1}$  & $4.381 \times 10^{-5}$ & $0.9753$ & $4.3 \times 10^{-3}$ \\
\ding{55}\, ST$^{\rm B}_8$\,    & $0.7$    & $0.078$  & $2.0 \times 10^{-3}$ & $0.311$ & $2.9 \times 10^{-1}$  & $4.740 \times 10^{-5}$ & $0.9800$ & $2.8 \times 10^{-3}$ \\
\hline\hline
\end{tabular}
\caption{Benchmark models for the sidetracked inflation model with the Starobinsky$+$quadratic potential~\eqref{eq:V_SQ} on the hyperbolic field space~\eqref{eq:metric_sidetracked}. Each series uses the same set of $L_{\rm fs}/L_{\rm crit}$ ratios~\eqref{eq:Lfs_scan}, which universally parametrizes the distance from the destabilization threshold~\eqref{eq:Lcrit}. The equivalent $\alpha$-attractor parameter is $\alpha = L_{\rm fs}^2/(3M_{\rm Pl}^2)$~\eqref{eq:Lfs_alpha}. All models are initialized at $(\phi_{S,i}, \phi_{Q,i}) = (6, 10^{-4})\,M_{\rm Pl}$ and evolved forward; the columns $\phi_{S,*}$ and $\phi_{Q,*}$ give the field values at the CMB pivot scale, $N_* = 55$ $e$-folds before the end of inflation. The Starobinsky-sector mass $m_S$ is fixed in each model by the amplitude normalization $\mathcal{P}_\mathcal{R}(k_*) = A_s$~\eqref{eq:As_normalization}. A check ($\checkmark$) or cross (\ding{55}) next to the model label indicates consistency with the joint Planck$+$BK18 constraints~\cite{Planck:2018jri, BICEP:2021xfz}, taking $n_s \in [0.959, 0.974]$ and $r < 0.036$ at 95\% CL.}
\label{tab:ST_runs}
\end{table}

The numerical results reveal a rich structure as the curvature is varied, as illustrated by the field-space trajectories in Figs.~\ref{fig:STA_trajectories}--\ref{fig:STB_trajectories} and the CMB observables in Table~\ref{tab:ST_runs}. In the \emph{stable regime} ($L_{\rm fs}/L_{\rm crit} \geq 5$), the trajectory remains close to the valley floor ($\phi_{Q,*} \ll L_{\rm fs}$) and the CMB observables are perturbatively close to the flat Starobinsky values ($n_s \simeq 0.966$, $r \simeq 0.004$). The field-space curvature introduces small but monotonic corrections: in Series~A, $r$ increases from $0.0036$ to $0.0059$ as $L_{\rm fs}/L_{\rm crit}$ decreases from $\infty$ to $5$, reflecting the growing geometric contribution to the effective mass~\eqref{eq:mu_s_valley}. In Series~B, $r$ instead \emph{decreases} from $0.0035$ to $0.0005$ over the same range, because the lighter $\phi_Q$ field allows the geometry to more efficiently redistribute power from adiabatic to entropic modes. All stable-regime models in both series are consistent with the Planck$+$BK18 constraints.

Near the \emph{destabilization threshold} ($L_{\rm fs}/L_{\rm crit} \simeq 1$--$2$), the entropic effective mass approaches and crosses zero, and the trajectory is driven away from the valley floor. This transition region produces disallowed CMB observables, with $n_s \simeq 1$ for the near-threshold models in Series~A (ST$^{\rm A}_6$--ST$^{\rm A}_7$) and $n_s$ outside the $2\sigma$ range in Series~B (ST$^{\rm B}_6$). Physically, this reflects strong superhorizon sourcing of the adiabatic perturbation $\mathcal{R}$ by the nearly tachyonic entropic mode, which induces pronounced scale dependence and, in the Series~A threshold runs, a blue scalar tilt. These models are excluded by CMB observations and define a \emph{forbidden band} in the $L_{\rm fs}$ parameter space.

In the \emph{sub-threshold regime} ($L_{\rm fs}/L_{\rm crit} = 0.7$), the behavior depends on the mass ratio. For Series~A ($m_Q \sim H_{*}$), the trajectory has settled onto a sidetracked solution by the pivot scale: $\phi_S$ is well off the Starobinsky plateau ($\phi_{S,*} \simeq 0.31\,M_{\rm Pl}$), the inflationary energy scale is dramatically reduced ($m_S \simeq 3 \times 10^{-7}\,M_{\rm Pl}$), and the tensor-to-scalar ratio at the pivot scale is negligible ($r \sim 10^{-7}$). For Series~B ($m_Q \ll H_{*}$), the sub-threshold models ST$^{\rm B}_7$--ST$^{\rm B}_8$ reach large $\phi_Q$ displacements at the pivot scale ($\phi_{Q,*} \simeq 0.3$--$0.4\,M_{\rm Pl}$) but fail the $n_s$ constraint at the $2\sigma$ level.

These results have important implications for gravitational particle production. The stable-regime models provide a controlled set of backgrounds in which field-space curvature modestly modifies the end-of-inflation dynamics and hence the GPP spectrum. The sidetracked model ST$^{\rm A}_8$, with its qualitatively different inflationary history, i.e., lower energy scale, a different oscillation pattern, and correlated post-inflationary motion in both field directions, represents a genuinely distinct background for particle production. Together with the flat Starobinsky$+$quadratic benchmarks of Sec.~\ref{subsec:starobinsky_quadratic}, these sidetracked models allow us to disentangle two effects on gravitational particle production:
\begin{enumerate}
\item \emph{Effect of initial conditions at fixed flat geometry.} The SQ$_1$--SQ$_{10}$ scan (Table~\ref{tab:SQ_runs}) varies the initial field displacement at $\mathcal{R}_{\rm fs}=0$, probing how the relative weighting of the Starobinsky and quadratic sectors modifies $a(t)$, $H(t)$, and hence the GPP spectrum.

\item \emph{Effect of field-space geometry at fixed potential and initial conditions.} The ST benchmarks vary $L_{\rm fs}$ at fixed $V_{\rm SQ}$ and fixed initial field values, isolating the role of field-space curvature in modifying the inflationary background, and in particular the post-inflationary transition, that drives gravitational production of $\chi$.
\end{enumerate}
We present the numerical results for the $\chi$ dark matter abundance in Sec.~\ref{sec:cgpp}.

\section{Cosmological Gravitational Particle Production}
\label{sec:cgpp}
In this section, we compute the gravitational production of a spectator scalar field $\chi$ during and after inflation, using the multifield backgrounds developed in Sec.~\ref{sec:benchmark_models}. The field $\chi$ interacts only gravitationally and serves as a dark matter candidate. We first review the general formalism of cosmological gravitational particle production (CGPP) following Refs.~\cite{Parker:1969au, Zeldovich:1971mw, Ford:1986sy, Chung:1998zb, Chung:1998ua, Kolb:2023ydq}, then discuss the role of the nonminimal coupling $\xi$ and its connection to fermionic production, and finally present the numerical spectra for the Starobinsky$+$quadratic and sidetracked benchmarks.

\subsection{Formalism}
\label{subsec:cgpp_formalism}
We consider a real scalar field $\chi$ of mass $m_\chi$ with action
\begin{equation}
S_\chi \;=\; \int d^4x\,\sqrt{-g}\left[\frac{1}{2}\,g^{\mu\nu}\partial_\mu\chi\,\partial_\nu\chi - \frac{1}{2}\left(m_\chi^2 + \xi R\right)\chi^2\right] ,
\label{eq:S_chi}
\end{equation}
where $R$ is the spacetime Ricci scalar and $\xi$ is a dimensionless nonminimal coupling.\footnote{Our sign convention $-\xi R$ in the action~\eqref{eq:S_chi} corresponds to $R > 0$ during de~Sitter inflation [Eq.~\eqref{eq:Ricci_scalar}].} The case $\xi = 0$ corresponds to \emph{minimal coupling}, while $\xi = 1/6$ corresponds to \emph{conformal coupling}, for which the massless theory is invariant under Weyl rescalings of the metric~\cite{Birrell:1982ix, Parker:2009uva}. From the perspective of effective field theory, the operator $\xi R\chi^2$ is the unique dimension-four scalar coupling to curvature consistent with the symmetries of the action, and should therefore be included on general grounds. Even if $\xi$ vanishes at tree level, radiative corrections generically induce a nonzero value~\cite{Callan:1970ze, Bunch:1980br, Bunch:1980bs, Birrell:1982ix, Odintsov:1990mt, Parker:2009uva}. For $\xi = 1/6$, particle production is driven entirely by the explicit breaking of conformal invariance through the mass term $m_\chi$, and the resulting mode equation is closely analogous to that of a massive spin-$1/2$ Dirac fermion minimally coupled to gravity~\cite{Kolb:2023ydq}. The conformally coupled scalar therefore provides a useful proxy for fermionic dark matter production, a connection we exploit below.

In the spatially flat FLRW spacetime $ds^2 = a^2(\eta)(d\eta^2 - |d\mathbf{x}|^2)$, where $\eta$ is conformal time, we define the rescaled field $X(\eta, \mathbf{x})= a(\eta) \chi(\eta, \mathbf{x})$ and decompose $X$ in Fourier modes
\begin{equation}
X(\eta, \mathbf{x}) \;=\; \int \frac{d^3\mathbf{k}}{(2\pi)^3}
\left(\hat a_{\mathbf{k}}\,\chi_k(\eta)\,e^{i\mathbf{k}\cdot\mathbf{x}}
+ \hat a_{\mathbf{k}}^\dagger\,\chi_k^*(\eta)\,e^{-i\mathbf{k}\cdot\mathbf{x}}\right) ,
\label{eq:chi_fourier}
\end{equation}
where we label modes by a comoving wavevector $\mathbf{k}$. Here the annihilation and creation operators, $\hat a_{\mathbf{k}}$ and $\hat a_{\mathbf{k}}^\dagger$, satisfy the usual canonical commutation relations. The mode functions $\chi_k(\eta)$ satisfy the equation of motion
\begin{equation}
\partial_\eta^2 \chi_k + \omega_k^2(\eta)\,\chi_k \;=\; 0 \,,
\label{eq:mode_eq_cgpp}
\end{equation}
with the time-dependent effective frequency
\begin{equation}
\omega_k^2(\eta) \;=\; k^2 + a^2 m_\chi^2 - (1-6\xi)\,\frac{\partial_\eta^2 a}{a} \,,
\label{eq:omega_k}
\end{equation}
where $k = |\mathbf{k}|$ is the comoving wavenumber. The last term can equivalently be written as $(\xi - 1/6)\,a^2 R$, using the Ricci scalar
\begin{equation}
R \;=\; \frac{6}{a^3}\,\partial_\eta^2 a \;=\; 6\left(\dot H + 2H^2\right) ,
\label{eq:Ricci_scalar}
\end{equation}
where dots denote derivatives with respect to cosmic time $t$ and $H = \dot a/a$. Equation~\eqref{eq:omega_k} makes clear that CGPP is driven by the two sources of conformal-symmetry breaking: the mass term $a^2 m_\chi^2$ and the gravitational coupling $(1 - 6\xi)\,\partial_\eta^2 a/a$. For a conformally coupled massless scalar ($\xi = 1/6$, $m_\chi = 0$), $\omega_k^2 = k^2$ is time-independent and no particles are produced~\cite{Parker:1969au, Birrell:1982ix}.

The particle number produced in each mode is encoded in the Bogoliubov coefficient $\beta_k$, which measures the projection of the evolved solution onto the negative-frequency basis at late times~\cite{Birrell:1982ix}. Given a solution to Eq.~\eqref{eq:mode_eq_cgpp} with the Bunch--Davies vacuum initial conditions
\begin{align}
\chi_k(\eta) &\;\xrightarrow{\eta \to -\infty}\; \sqrt{\frac{1}{2k}}\,e^{-ik\eta} \,,
\nonumber\\[4pt]
\partial_\eta \chi_k(\eta) &\;\xrightarrow{\eta \to -\infty}\; -i\sqrt{\frac{k}{2}}\,e^{-ik\eta} \,,
\label{eq:BD_init}
\end{align}
the occupation number at late times is obtained from
\begin{equation}
|\beta_k|^2 \;=\; \frac{\omega_k}{2}\,|\chi_k|^2 + \frac{|\partial_\eta \chi_k|^2}{2\omega_k} - \frac{1}{2} \,.
\label{eq:beta_k}
\end{equation}
The comoving particle number density is then given by
\begin{equation}
n_\chi\,a^3 \;=\; \int d\log k\;\,a^3 n_k \qquad\text{where}\qquad a^3 n_k \;\equiv\; \frac{k^3}{2\pi^2}\,|\beta_k|^2 \,.
\label{eq:spectrum_cgpp}
\end{equation}
The spectrum $n_k$ quantifies the comoving number density of $\chi$ particles produced at each comoving wavenumber $k$.

In practice, we evolve Eq.~\eqref{eq:mode_eq_cgpp} numerically for each $k$, starting from the Bunch--Davies condition~\eqref{eq:BD_init} at an early conformal time when $k/(aH)\gg 1$, through the end of inflation and into the post-inflationary oscillation phase. The Bogoliubov coefficient $|\beta_k|^2$ is then extracted from Eq.~\eqref{eq:beta_k} once the mode evolution has become adiabatic and the occupation number has converged to its asymptotic value.

\subsection{CGPP in Multifield Inflation}
\label{cgpp:multifield}
Before presenting the numerical spectra, it is useful to identify the qualitative features that distinguish gravitational particle production in multifield inflation from the better-understood single-field case. The key observable is the spectrum $n_k$, which is shaped by the time-dependent effective frequency $\omega_k^2(\eta)$ in Eq.~\eqref{eq:omega_k}. The background enters $\omega_k^2$ through the scale factor $a(\eta)$ and, for $\xi \neq 1/6$, through the Ricci scalar $R(\eta)$. Both quantities are determined by the post-inflationary dynamics of the inflaton sector, which can differ qualitatively between single-field and multifield models.

In terms of the background fields, the Ricci scalar~\eqref{eq:Ricci_scalar} can be written as
\begin{equation}
R \;=\; \frac{4V - \dot\sigma^2}{M_{\rm Pl}^2} \,,
\label{eq:R_background}
\end{equation}
where $\dot\sigma^2 = G_{IJ}\dot\phi^I\dot\phi^J$ is the total kinetic energy in the inflaton sector. During slow-roll inflation, the kinetic energy is negligible and $R \simeq 4V/M_{\rm Pl}^2 \simeq 12H^2$. At the end of inflation, the fields begin to oscillate about the potential minimum and $R$ oscillates in turn.

In a \emph{single-field} model with a quadratic potential $V = m^2\phi^2/2$, the inflaton oscillates coherently at frequency $m$ after inflation, the time-averaged equation of state is $\langle w \rangle = 0$ (matter-like), and $R$ oscillates at frequency $2m$ around a slowly decaying mean $\langle R\rangle = 3H^2$, crossing zero twice per inflaton cycle. The CGPP spectrum reflects these oscillations through the effective frequency $\omega_k^2(\eta)$ [Eq.~\eqref{eq:omega_k}], but the dominant mechanism depends on the nonminimal coupling $\xi$. For minimal coupling ($\xi = 0$), the curvature term $(1-6\xi)\,\partial_\eta^2 a/a$ in $\omega_k^2$ is proportional to $a^2 R$, so each zero crossing of $R$ produces a burst of $\chi$ particles, and the regular spacing of these crossings gives rise to the well-known staircase-like interference pattern in the UV tail of the CGPP spectrum~\cite{Basso:2022tpd, Ema:2018ucl, Kolb:2023ydq}. For conformal coupling ($\xi = 1/6$), the curvature term vanishes identically and production is instead driven by the oscillating mass term $a^2(\eta)\,m_\chi^2$, which inherits the inflaton matter-dominated oscillations through the Friedmann equation.

In a \emph{two-field} model, both $\phi_S$ and $\phi_Q$ oscillate after inflation, generically with different frequencies $m_S$ and $m_Q$. For the flat field-space benchmarks of Sec.~\ref{subsec:starobinsky_quadratic}, the potential is separable and the two fields oscillate independently, each contributing to the Ricci scalar~\eqref{eq:R_background} primarily at frequencies $2m_S$ and $2m_Q$. The superposition of these oscillating contributions produces a beating pattern in $R(\eta)$ with an envelope that modulates on the timescale $\sim \pi/|m_S-m_Q|$. The resulting zero crossings of $R$ are no longer equally spaced: when the two contributions are in phase, $R$ oscillates with enhanced amplitude, while destructive interference can temporarily suppress the oscillation and delay zero crossings. This beating directly imprints on the CGPP spectrum as a modification of the UV interference features relative to the single-field case.
 
More generally, in multifield models the post-inflationary trajectory need not oscillate back and forth along a single line in field space. Instead, it can follow a curved, multi-frequency path toward the potential minimum at $(\phi_S,\phi_Q)=(0,0)$, as illustrated by the SQ and ST models in Figs.~\ref{fig:SQ_trajectories}--\ref{fig:STB_trajectories}. The evolving relative phase between the two fields produces a time-dependent effective equation of state $w(t)$ that differs from the single-field matter-like average $\langle w \rangle = 0$. Since the CGPP spectrum at UV scales ($k \gtrsim k_{\rm e} = a_{\rm e} H_{\rm e}$, where $a_{\rm e}$ and $H_{\rm e}$ are the scale factor and Hubble parameter at the end of inflation) is sensitive to the detailed time dependence of $a(\eta)$ during the first several post-inflationary oscillation cycles, these multifield effects leave characteristic signatures in $n_k$.

\paragraph{Infrared versus ultraviolet regimes.}
The CGPP spectrum separates naturally into two regimes:
\begin{itemize}
\item \emph{Infrared} ($k \ll k_{\rm e}$): These modes exit the horizon well before the end of inflation and are amplified on superhorizon scales. For $\xi = 0$, the curvature coupling in Eq.~\eqref{eq:omega_k} provides an effective mass contribution of order $H$ during quasi-de~Sitter inflation. In the usual light-spectator limit ($m_\chi \ll H_{*}$), the superhorizon spectrum is approximately scale-invariant, with a slight red tilt controlled by $m_\chi^2/H_{*}^2$, whereas for heavier fields ($m_\chi \gtrsim H_{*}$) it is roughly exponentially suppressed~\cite{Kolb:2023ydq}. Although the IR plateau is primarily set by the inflationary Hubble scale, it is not universal across multifield benchmarks: its height depends on the slow-roll evolution $\varepsilon(N)$ and on the details of the inflation to oscillation transition. As shown in Fig.~\ref{fig:cgpp_SQ}, the pure quadratic limit (SQ$_1$) produces an IR plateau several times higher than the pure Starobinsky limit (SQ$_{10}$), reflecting the sharper end of inflation transition and the more abrupt onset of post-inflationary oscillations in the quadratic case. For $\xi = 1/6$, the curvature coupling vanishes and the IR spectrum is strongly suppressed, since particle production in this regime is driven only by the mass term $a^2m_\chi^2$.

\item \emph{Ultraviolet} ($k \gtrsim k_{\rm e}$): These modes are near the horizon at the end of inflation and are maximally sensitive to the transition from inflation to the oscillatory phase. The sharpness of this transition, the initial amplitude and phase of the post-inflationary oscillations, and the subsequent beating pattern in $R$ all leave imprints on $n_k$ at these scales. Modes with $k > k_{\rm e}$ that never exit the horizon can still be produced by the nonadiabatic evolution of $\omega_k$ during the post-inflationary oscillations; this gives rise to the interference features visible in the UV tails of the spectra, which can be understood as quantum interference between successive particle-production events associated with repeated zero crossings of $R$~\cite{Basso:2022tpd}.
\end{itemize}

\paragraph{Minimally coupled versus conformally coupled production.}
The distinction between $\xi = 0$ and $\xi = 1/6$ acquires 
additional significance in the multifield context. For $\xi = 0$, 
the curvature coupling in Eq.~\eqref{eq:omega_k} ties $\chi$ 
directly to the Ricci scalar $R(\eta)$, and therefore to the full 
oscillation pattern of the inflaton sector after inflation. This is 
the dominant production channel for light spectators 
($m_\chi \ll H_{\rm e}$), and the resulting spectra carry detailed 
information about the post-inflationary trajectory---including the 
multifrequency beating structure discussed above. For $\xi = 1/6$, 
the Ricci scalar drops out entirely, and $\chi$ couples to the 
background only through $a^2(\eta)\,m_\chi^2$. The scale factor 
$a(\eta)$ is an integrated quantity (determined by the Friedmann 
equation) and is therefore less sensitive to the oscillatory details 
than $R(\eta)$ itself. As a result, the $\xi = 1/6$ spectra show 
less variation across the multifield benchmarks than the $\xi = 0$ 
spectra, as we confirm numerically below.

\paragraph{Multifield effects on the equation of state.}
A useful diagnostic of the post-inflationary dynamics is the effective 
equation of state $w \equiv p/\rho = (\dot\sigma^2/2 - V)/
(\dot\sigma^2/2 + V)$. For a single field oscillating in a quadratic 
potential, $\langle w \rangle = 0$ at all amplitudes. The Starobinsky 
potential is quadratic near the minimum but asymmetric at large field 
values (flat plateau on one side), so the time-averaged equation of 
state deviates from zero during the first several oscillation cycles 
when the amplitude is still large, before relaxing to $\langle w \rangle \to 0$ as the field settles into the quadratic regime~\cite{Turner:1983he, Martin:2010kz, Garcia:2020eof, Garcia:2020wiy}. In two-field models, the total energy density receives independent oscillating contributions from both $\phi_S$ and $\phi_Q$, each with a different frequency ($m_S$ and $m_Q$ respectively). The Ricci scalar inherits oscillations at $2m_S$ and $2m_Q$, and their superposition produces a beating pattern with an envelope that modulates on the timescale $\sim \pi/|m_S - m_Q|$. For the mass ratio $m_S/m_Q = 1.92$, this corresponds to a beat period of order a few $R$-oscillation cycles, leading to a rapid modulation of the post-inflationary dynamics that is visible in the CGPP spectra as enhanced UV structure relative to the single-field limits.

For the sidetracked benchmarks, the curved field-space metric  introduces genuine coupling between the two oscillation modes through  the Christoffel symbols in Eqs.~\eqref{eq:bg1_sidetracked}--\eqref{eq:bg2_sidetracked}: the centrifugal term mixes the two fields during the post-inflationary phase, and the metric factor $e^{2\phi_Q/L_{\rm fs}}$ modulates the contribution of $\phi_S$ to the kinetic energy. Unlike the flat field-space case, where the two fields oscillate independently and interact only through the shared Hubble friction, the curved geometry induces direct energy exchange between the two sectors, further modifying the post-inflationary evolution of $R(\eta)$ and $a(\eta)$, and hence the CGPP spectrum.

\subsection{CGPP in Starobinsky$+$Quadratic Inflation}
\label{subsec:cgpp_SQ}
\begin{figure}[t]
\centering
\includegraphics[width=\textwidth]{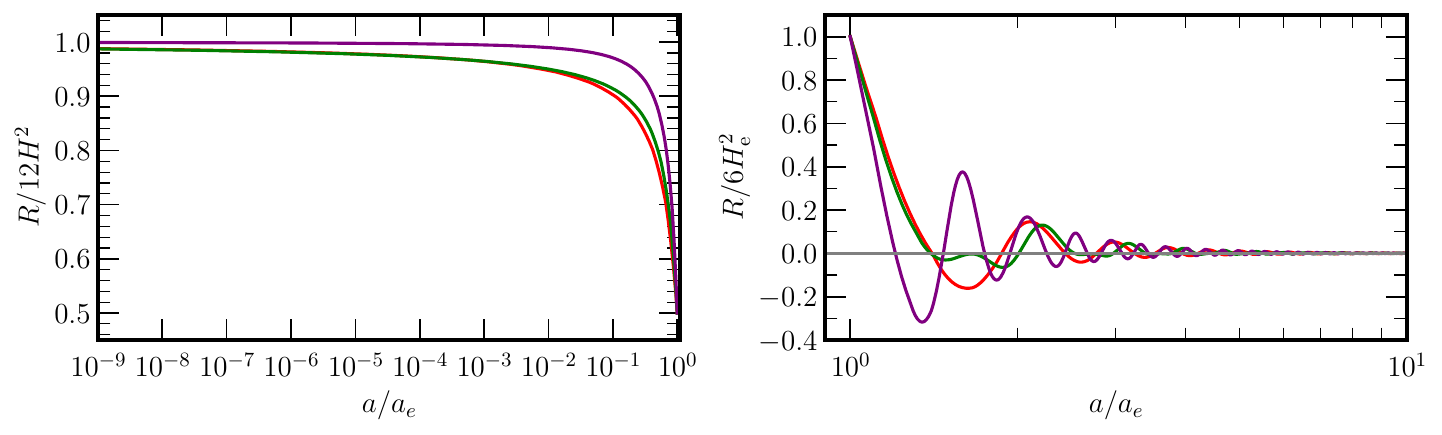}
\caption{Evolution of the Ricci scalar for the benchmarks SQ$_1$ (pure quadratic, red), SQ$_5$ (intermediate, green), and SQ$_{10}$ (pure Starobinsky, purple). Left panel: $R/(12H^2)$ as a function of $a/a_{\rm e}$ during inflation. In the quasi-de~Sitter phase $R/(12H^2) \simeq 1$. The departure from unity reflects the growth of the slow-roll parameter $\varepsilon$. Right panel: $R/(6H_{\rm e}^2)$ as a function of $a/a_{\rm e}$ during the post-inflationary phase. The three benchmarks exhibit visibly different oscillation patterns: the pure quadratic and pure Starobinsky limits show distinct waveforms and amplitudes, while the intermediate run displays a characteristic beating pattern associated with the superposition of oscillations from the two inflaton-sector fields.}
\label{fig:ricciSQ}
\end{figure}

Before presenting the CGPP spectra, it is instructive to examine the evolution of the Ricci scalar $R$ across the SQ benchmarks, since for $\xi \neq 1/6$ it is $R(\eta)$ that predominantly drives particle production through the $(1 - 6\xi)\,\partial_\eta^2 a/a$ term in Eq.~\eqref{eq:omega_k}. Figure~\ref{fig:ricciSQ} shows $R$ normalized to $12H^2$ (left panel) during inflation, and normalized to $6H_{\rm e}^2$ (right panel) during the post-inflationary phase, for three representative benchmarks: SQ$_1$ (pure quadratic), SQ$_5$ (intermediate), and SQ$_{10}$ (pure Starobinsky).

During slow-roll inflation, $R/(12H^2) = (2 - \varepsilon)/2 \simeq 1 - \varepsilon/2$ is close to unity for all models, deviating only as $\varepsilon$ grows toward the end of inflation. The pure quadratic run (SQ$_1$) departs from unity earliest because the slow-roll parameter $\varepsilon = 1/(2N)$ grows more steeply than the Starobinsky value $\varepsilon = 3/(4N^2)$.

After inflation (right panel), the Ricci scalar oscillates with model-dependent frequency, amplitude, and waveform. For the pure quadratic case (SQ$_1$), the inflaton oscillates symmetrically in a parabolic potential at frequency $m_Q$, and $R$ oscillates at $2m_Q$ with a regular, sinusoidal waveform and slowly decaying envelope. For the pure Starobinsky case (SQ$_{10}$), the situation is qualitatively different because the Starobinsky potential near the minimum contains a cubic self-interaction. Expanding~\eqref{eq:V_SQ} about $\phi_S = 0$ gives
\begin{equation}
V_{\rm S}(\phi_S) \;\simeq\; \frac{1}{2}\,m_S^2\,\phi_S^2 \left(1 - \sqrt{\frac{2}{3}}\,\frac{\phi_S}{M_{\rm Pl}} + \cdots\right) ,
\label{eq:V_cubic}
\end{equation}
where the cubic term $\propto \phi_S^3/M_{\rm Pl}$ breaks the $\phi_S \to -\phi_S$ symmetry of the quadratic approximation. Physically, the potential is shallower for $\phi_S>0$ (toward the plateau) and steeper for $\phi_S<0$, so the post-inflationary oscillation is asymmetric: the inflaton spends more time on the plateau side and less time on the steep side during each cycle. This asymmetry distorts the waveform of $R$ away from a pure sinusoid, introducing higher harmonics and making the positive and negative half-cycles of $R$ unequal in both amplitude and duration. In the inflaton-scattering interpretation of Ref.~\cite{Basso:2022tpd}, where the oscillating condensate annihilates via graviton exchange into $\chi$ pairs, the cubic self-interaction introduces additional scattering channels ($\phi\phi\phi \to \chi\chi$) with distinct kinematics, and the asymmetry between successive half-cycles modifies the quantum interference pattern in the UV spectrum.

For the intermediate run SQ$_5$ (green), both $\phi_S$ and $\phi_Q$ oscillate after inflation, and the Ricci scalar inherits contributions primarily at both frequencies $2m_S$ and $2m_Q$. The superposition of these two independently oscillating contributions produces the \emph{beating pattern} visible in the right panel of Fig.~\ref{fig:ricciSQ}, where the amplitude of $R$ modulates on a timescale $\sim \pi/|m_S - m_Q|$.

\begin{figure}[t]
\centering
\includegraphics[width=\textwidth]{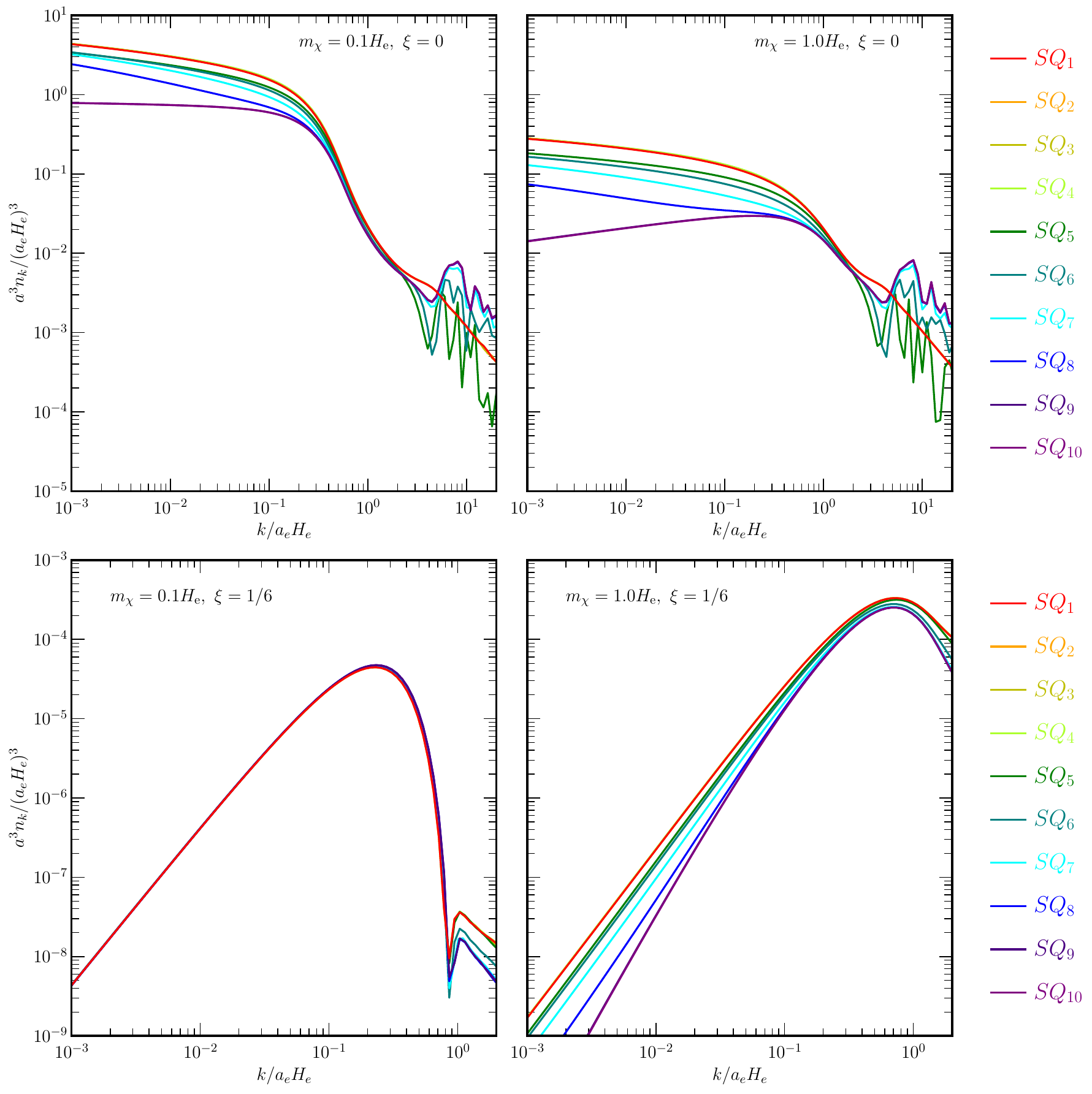}
\caption{Gravitational particle production spectra $n_k/(a_{\rm e}H_{\rm e})^3$ for the Starobinsky$+$quadratic benchmarks SQ$_1$--SQ$_{10}$ (Table~\ref{tab:SQ_runs}), plotted as a function of $k/(a_{\rm e}H_{\rm e})$. Top left: $m_\chi = 0.1\,H_{\rm e}$, $\xi = 0$. Top right: $m_\chi = H_{\rm e}$, $\xi = 0$. Bottom left: $m_\chi = 0.1\,H_{\rm e}$, $\xi = 1/6$. Bottom right: $m_\chi = H_{\rm e}$, $\xi = 1/6$. The color coding runs from SQ$_1$ (red, pure quadratic) to SQ$_{10}$ (purple, pure Starobinsky).}
\label{fig:cgpp_SQ}
\end{figure}

We compute the CGPP spectrum $n_k$ for all SQ benchmarks of Table~\ref{tab:SQ_runs}, for two representative values of the spectator mass ($m_\chi = 0.1\,H_{\rm e}$ and $m_\chi = H_{\rm e}$) and two values of the nonminimal coupling ($\xi = 0$ and $\xi = 1/6$). The results are shown in Fig.~\ref{fig:cgpp_SQ}.

\emph{Minimally coupled case} ($\xi = 0$, top panels). The spectra exhibit a broad plateau at infrared scales ($k \ll k_{\rm e} \equiv a_{\rm e}H_{\rm e}$), followed by a rapid falloff modulated by oscillatory interference features at $k \gtrsim k_{\rm e}$. The IR plateau arises because long-wavelength modes that exit the horizon during inflation are superhorizon-amplified by the curvature term $(1-6\xi)\,\partial_\eta^2 a/a$ in Eq.~\eqref{eq:omega_k}. In the exact de~Sitter limit, the superhorizon solution is controlled by the Hankel index $\nu = \sqrt{9/4 - m_\chi^2/H_{*}^2}$~\cite{Kolb:2023ydq}, so the spectrum acquires a tilt set by $m_\chi/H_{*}$: for $m_\chi \ll H_{*}$, the plateau is nearly flat, while for $m_\chi = H_{\rm e}$ (top right panel) a visible red tilt develops.

The height of the IR plateau is \emph{not} universal across benchmarks. The pure quadratic run SQ$_1$ (red) yields the highest plateau, roughly three to five times above the pure Starobinsky run SQ$_{10}$ (purple). This hierarchy reflects two effects: the different slow-roll histories $\varepsilon(N)$, which modify the superhorizon amplification, and the different sharpness of the inflation to oscillation transition, which affects the final occupation number as the modes evolve into the post-inflationary adiabatic regime. The Starobinsky-dominated runs SQ$_8$--SQ$_{10}$ produce nearly identical IR spectra, since these benchmarks share the same slow-roll evolution on the Starobinsky plateau and differ only in their subdominant $\phi_Q$ component. The spectra fan out only toward UV scales ($k \gtrsim 0.1\,k_{\rm e}$), where the post-inflationary oscillation pattern becomes important.

At ultraviolet scales ($k \gtrsim k_{\rm e}$), the spectra develop oscillatory interference features whose structure depends sensitively on the inflaton dynamics. For the quadratic-dominated runs (SQ$_1$--SQ$_3$), the interference pattern is rapid and extends to high $k/k_{\rm e}$, reflecting the large-amplitude, high-frequency oscillations of $R$ visible in Fig.~\ref{fig:ricciSQ}. Each zero crossing of $R$ produces a burst of $\chi$ particles, and quantum interference between successive bursts generates the characteristic oscillatory structure in $n_k$~\cite{Basso:2022tpd}. For the pure Starobinsky run (SQ$_{10}$), the UV tail is modified by the cubic self-interaction~\eqref{eq:V_cubic}: the asymmetric oscillation waveform produces unequal particle-production amplitudes across successive half-cycles, breaking the regularity of the interference pattern and introducing additional modulation~\cite{Basso:2022tpd, Kolb:2023ydq}. For intermediate runs such as SQ$_5$, the beating pattern in $R$ from the two-frequency superposition (Fig.~\ref{fig:ricciSQ}, right panel) translates into a modulated UV interference pattern with amplitude variations on the beat timescale.

\emph{Conformally coupled case} ($\xi = 1/6$, bottom panels). The curvature coupling $(1-6\xi)\,\partial_\eta^2 a/a$ vanishes identically, so the superhorizon amplification mechanism is absent. The spectra therefore peak near $k \sim k_{\rm e}$ and are suppressed by several orders of magnitude relative to the minimally coupled case, consistent with the fact that the only source of conformal-symmetry breaking is the mass term $a^2 m_\chi^2$. As emphasized in Sec.~\ref{subsec:cgpp_formalism}, these spectra are also closely analogous to those governing the gravitational production of massive spin-$1/2$ fermions, up to the appropriate spin degeneracy factor for Dirac fermions.

The ordering across benchmarks is particularly striking for $\xi = 1/6$: the pure quadratic run SQ$_1$ (red) produces a spectrum two to three orders of magnitude above the Starobinsky-dominated runs (SQ$_8$--SQ$_{10}$), which cluster tightly together. This is because the conformal spectrum is controlled by the post-inflationary expansion history encoded in $a^2(\eta)m_\chi^2$. The quadratic potential supports large-amplitude coherent oscillations with $\langle w \rangle = 0$ from the outset, leading to stronger nonadiabatic evolution and hence more efficient particle production. The Starobinsky potential, by contrast, yields a milder and more asymmetric post-inflationary evolution during the first several cycles, resulting in substantially weaker production.

Comparing the light ($m_\chi = 0.1\,H_{\rm e}$, bottom left) and heavy ($m_\chi = H_{\rm e}$, bottom right) conformally coupled cases, the heavier spectator yields a larger spectrum in these benchmarks because the explicit conformal symmetry breaking term $a^2 m_\chi^2$ is larger, leading to stronger nonadiabatic evolution of the mode functions. This is opposite to the minimally coupled case, where heavier particles are strongly suppressed because the inflationary superhorizon amplification becomes less efficient as $m_\chi/H$ increases.

\subsection{CGPP in Sidetracked Inflation}
\label{subsec:cgpp_ST}
\begin{figure}[t]
\centering
\includegraphics[width=\textwidth]{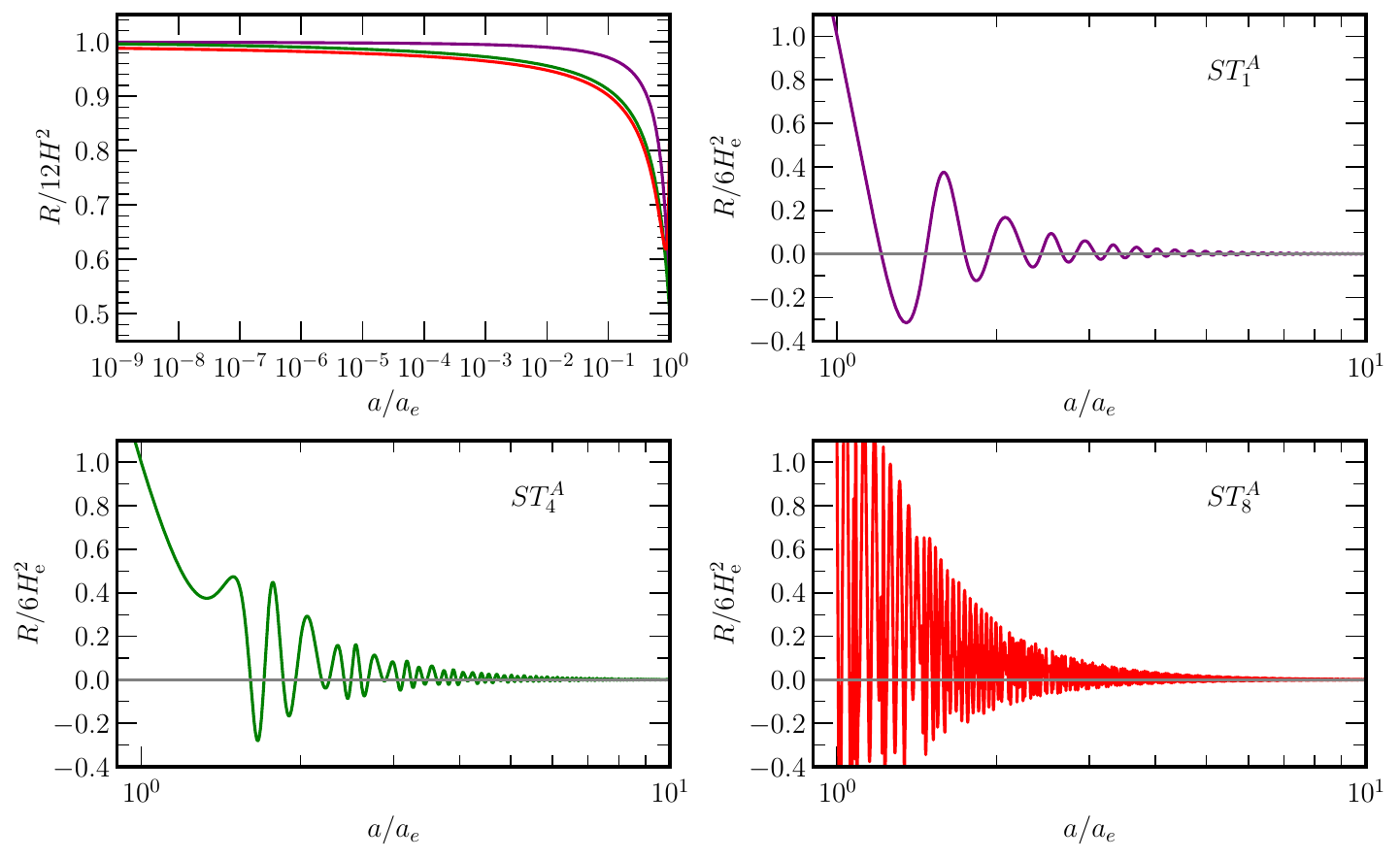}
\caption{Ricci scalar evolution for the sidetracked benchmarks of Series~A ($m_S/m_Q = 1.92$). Top left: $R/(12H^2)$ during inflation for ST$^{\rm A}_1$ (purple, flat limit), ST$^{\rm A}_4$ (green, $L_{\rm fs}/L_{\rm crit} = 10$), and ST$^{\rm A}_8$ (red, $L_{\rm fs}/L_{\rm crit} = 0.7$). Remaining panels: $R/(6H_{\rm e}^2)$ during the post-inflationary phase for each run individually. The flat limit (top right) reproduces the gentle Starobinsky oscillation pattern. The stable-regime run (bottom left) shows enhanced oscillation frequency from the geometric coupling. The sub-threshold run (bottom right) displays very rapid, high-frequency oscillations characteristic of the sidetracked regime.}
\label{fig:ricciSTA}
\end{figure}

\begin{figure}[t]
\centering
\includegraphics[width=\textwidth]{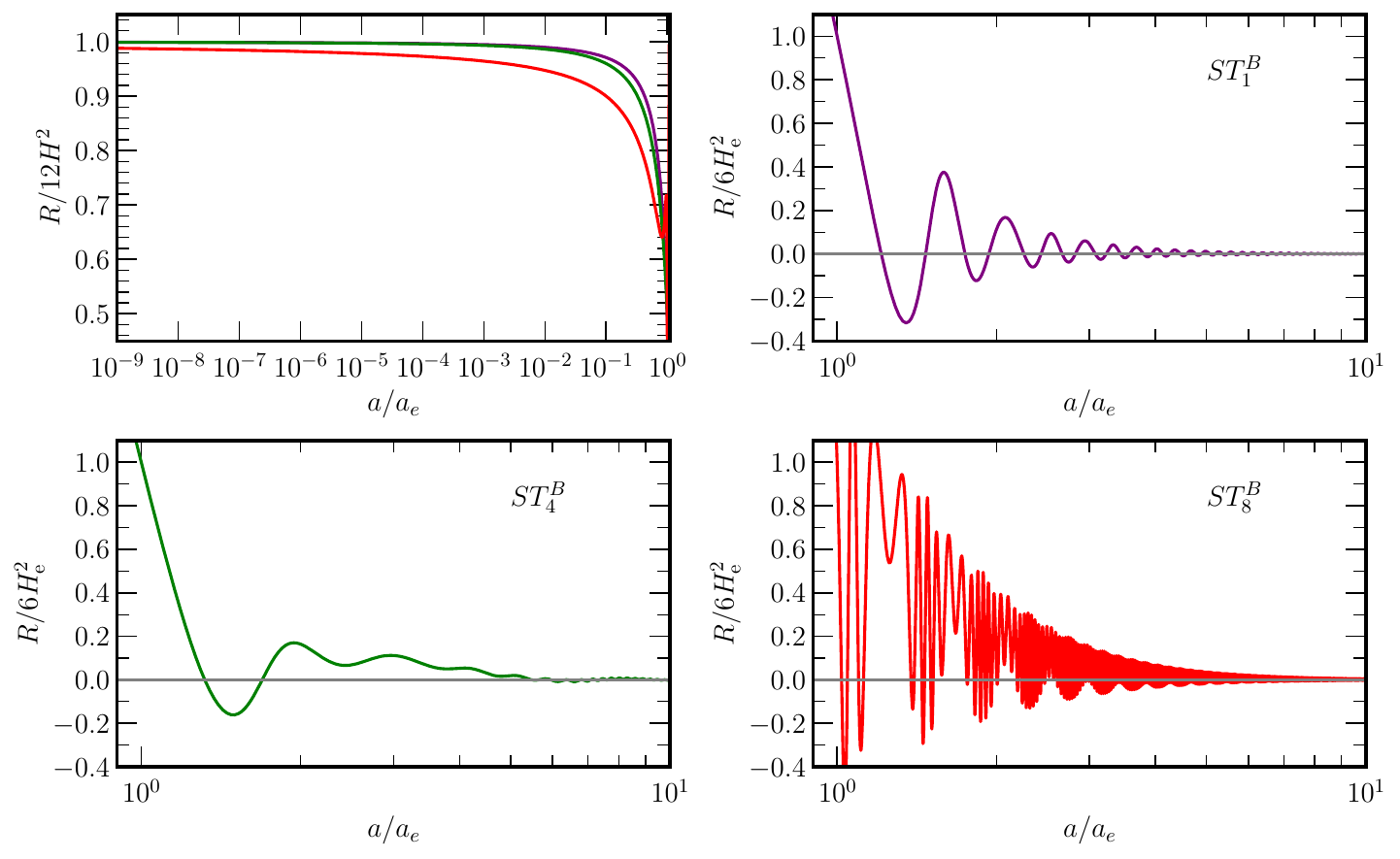}
\caption{Same as Fig.~\ref{fig:ricciSTA} but for Series~B ($m_S/m_Q = 10$). The lighter $\phi_Q$ field amplifies the effects of the curved field-space geometry: the stable-regime run ST$^{\rm B}_4$ (bottom left) shows a slowly damped oscillatory evolution of the Ricci scalar, while the sub-threshold run ST$^{\rm B}_8$ (bottom right) exhibits much more rapid oscillations whose amplitude decays gradually after inflation.}
\label{fig:ricciSTB}
\end{figure}

The post-inflationary evolution of $R$ in the sidetracked benchmarks differs qualitatively from the flat Starobinsky$+$quadratic case, as shown in Figs.~\ref{fig:ricciSTA} and~\ref{fig:ricciSTB}. Each figure displays three representative runs: the flat-limit benchmark ST$_1$ (purple), the stable-regime benchmark ST$_4$ ($L_{\rm fs}/L_{\rm crit} = 10$, green), and the sub-threshold benchmark ST$_8$ ($L_{\rm fs}/L_{\rm crit} = 0.7$, red).

During inflation (top-left panels), the normalized Ricci scalar $R/(12H^2)$ remains close to unity for all runs, as in the SQ case. The departure from unity occurs somewhat earlier for the sub-threshold runs, because the displacement of $\phi_Q$ from the valley floor modifies $\varepsilon$ and shortens the remaining duration of inflation. This effect is more pronounced in Series~B, where the lighter $\phi_Q$ field is more easily displaced.

The post-inflationary oscillations (remaining panels) reveal dramatic differences across the three regimes. The flat-limit run ST$_1$ (purple, top-right panels) reproduces the familiar gentle, low-frequency Starobinsky oscillation pattern. The stable-regime run ST$_4$ (green, bottom-left panels) already displays markedly different behavior: although the curved field-space geometry is not strong enough to trigger sidetracking during inflation, it nevertheless modifies the post-inflationary trajectory in the $(\phi_S,\phi_Q)$ plane. Both fields contribute to the kinetic energy $\dot\sigma^2$ in Eq.~\eqref{eq:R_background}, and the Christoffel symbols in Eqs.~\eqref{eq:bg1_sidetracked}--\eqref{eq:bg2_sidetracked} induce genuine coupling between the two oscillation modes, introducing additional modulation beyond the independent superposition seen in the flat SQ benchmarks. This effect is even more striking in Series~B (Fig.~\ref{fig:ricciSTB}), where the lighter $\phi_Q$ field oscillates more slowly and the geometric coupling generates richer high-frequency structure in $R$.

The sub-threshold run ST$_8$ (red, bottom-right panels) exhibits the most dramatic behavior: $R$ oscillates with very high frequency and large amplitude immediately after inflation. This reflects the correlated post-inflationary motion of both fields on the sidetracked solution, where the trajectory follows a strongly curved path toward the origin in the $(\phi_S,\phi_Q)$ plane (cf.\ Figs.~\ref{fig:STA_trajectories} and~\ref{fig:STB_trajectories}) while both fields oscillate simultaneously. The metric factor $e^{2\phi_Q/L_{\rm fs}}$ in the kinetic term modulates the effective oscillation frequency: when $\phi_Q$ is displaced, the $\phi_S$ contribution to the kinetic energy is amplified by this factor, producing rapid variations in $\dot\sigma^2$ and hence in $R$. The large number of zero crossings during the early post-inflationary phase is therefore expected to enhance UV particle production significantly relative to the flat-limit case.

The spectator field $\chi$ couples only to gravity and does not interact directly with the inflaton field-space metric $G_{IJ}$. In our setup, the effect of field-space curvature on CGPP therefore enters only through the background quantities $a(\eta)$ and $R(\eta)$ appearing in the mode equation~\eqref{eq:mode_eq_cgpp}: the curved geometry modifies the inflaton dynamics, which in turn modifies $a(\eta)$ and $R(\eta)$, and thereby the $\chi$ spectrum.

The mechanism by which negative field-space curvature enhances post-inflationary $R$ oscillations can be understood as follows. In the flat Starobinsky limit (ST$_1$), the post-inflationary dynamics is effectively single-field: $\phi_Q$ is either frozen at zero or rapidly damped, and $\phi_S$ oscillates gently in the asymmetric Starobinsky potential. The Ricci scalar, given by Eq.~\eqref{eq:R_background}, oscillates with modest amplitude because $\dot\sigma^2 \simeq \dot\phi_S^2$ and the Starobinsky potential is nearly quadratic near the minimum. When field-space curvature is turned on, the Christoffel symbols~\eqref{eq:christoffel_sidetracked} act as a geometric pump: the centrifugal term $-(e^{2\phi_Q/L_{\rm fs}}/L_{\rm fs})\dot\phi_S^2$ in the $\phi_Q$ equation~\eqref{eq:bg2_sidetracked} sources $\phi_Q$ oscillations even when $\phi_Q$ is initially near zero. Once $\phi_Q$ is excited, the field-space speed
\begin{equation}
\dot\sigma^2 = e^{2\phi_Q/L_{\rm fs}}\dot\phi_S^2 + \dot\phi_Q^2
\end{equation}
[Eq.~\eqref{eq:sigmadot_sidetracked}] receives an amplified contribution from the $\phi_S$ sector through the metric factor $e^{2\phi_Q/L_{\rm fs}}$. For a displacement $\phi_Q \sim L_{\rm fs}$, this factor is already of order $e^2 \simeq 7$, substantially increasing the amplitude and frequency content of the oscillations in $\dot\sigma^2$ and hence in $R$.

\begin{figure}[t]
\centering
\includegraphics[width=\textwidth]{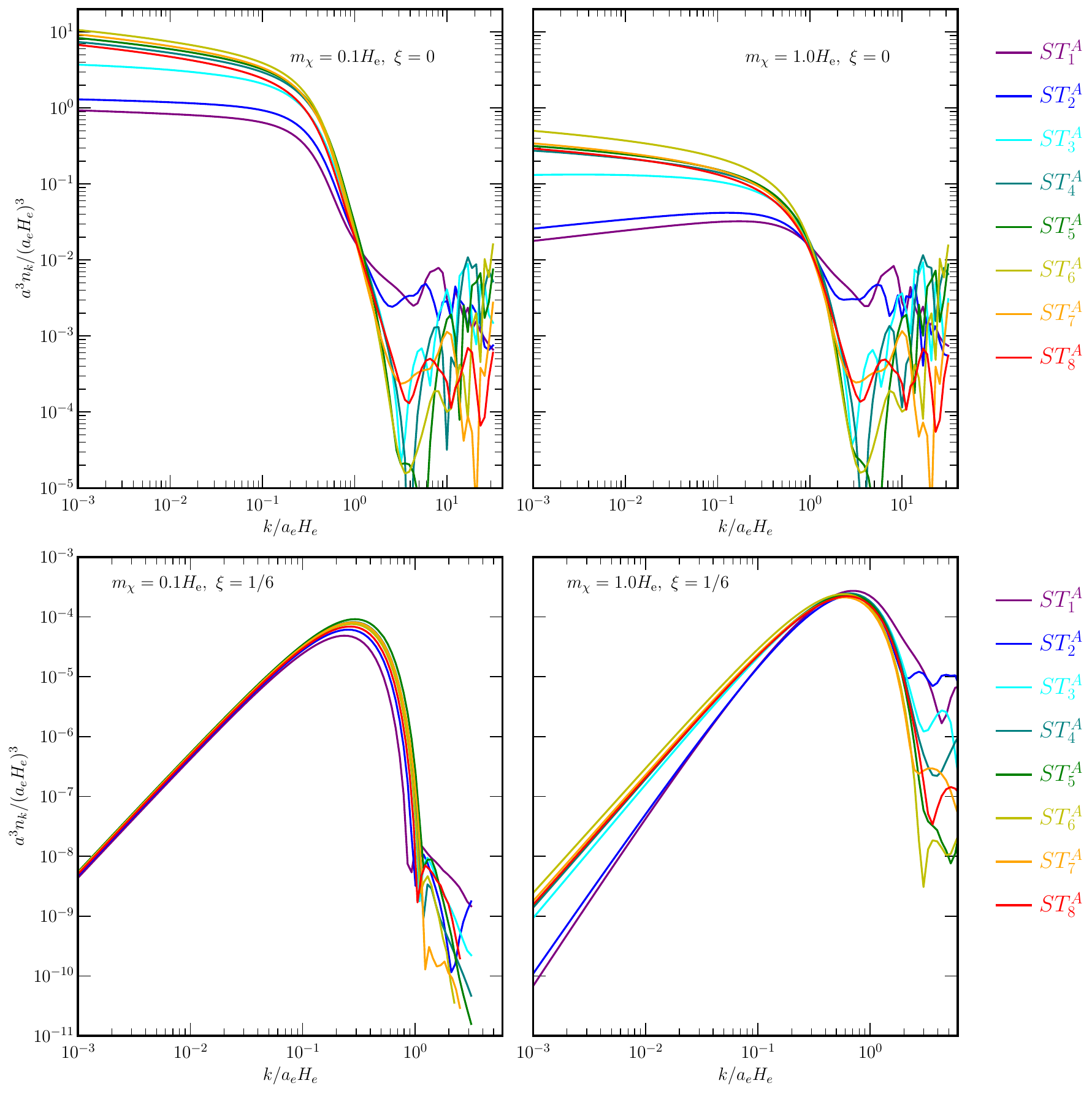}
\caption{Gravitational particle production spectra $a^3 n_k/(a_{\rm e}H_{\rm e})^3$ for the sidetracked benchmarks of Series~A ($m_S/m_Q = 1.92$, Table~\ref{tab:ST_runs}), plotted as a function of $k/(a_{\rm e}H_{\rm e})$. Top left: $m_\chi = 0.1\,H_{\rm e}$, $\xi = 0$. Top right: $m_\chi = H_{\rm e}$, $\xi = 0$. Bottom left: $m_\chi = 0.1\,H_{\rm e}$, $\xi = 1/6$. Bottom right: $m_\chi = H_{\rm e}$, $\xi = 1/6$. The color coding runs from ST$^{\rm A}_1$ (purple, flat limit) to ST$^{\rm A}_8$ (red, most strongly sidetracked benchmark in the scan).}
\label{fig:cgpp_STA}
\end{figure}

\begin{figure}[t]
\centering
\includegraphics[width=\textwidth]{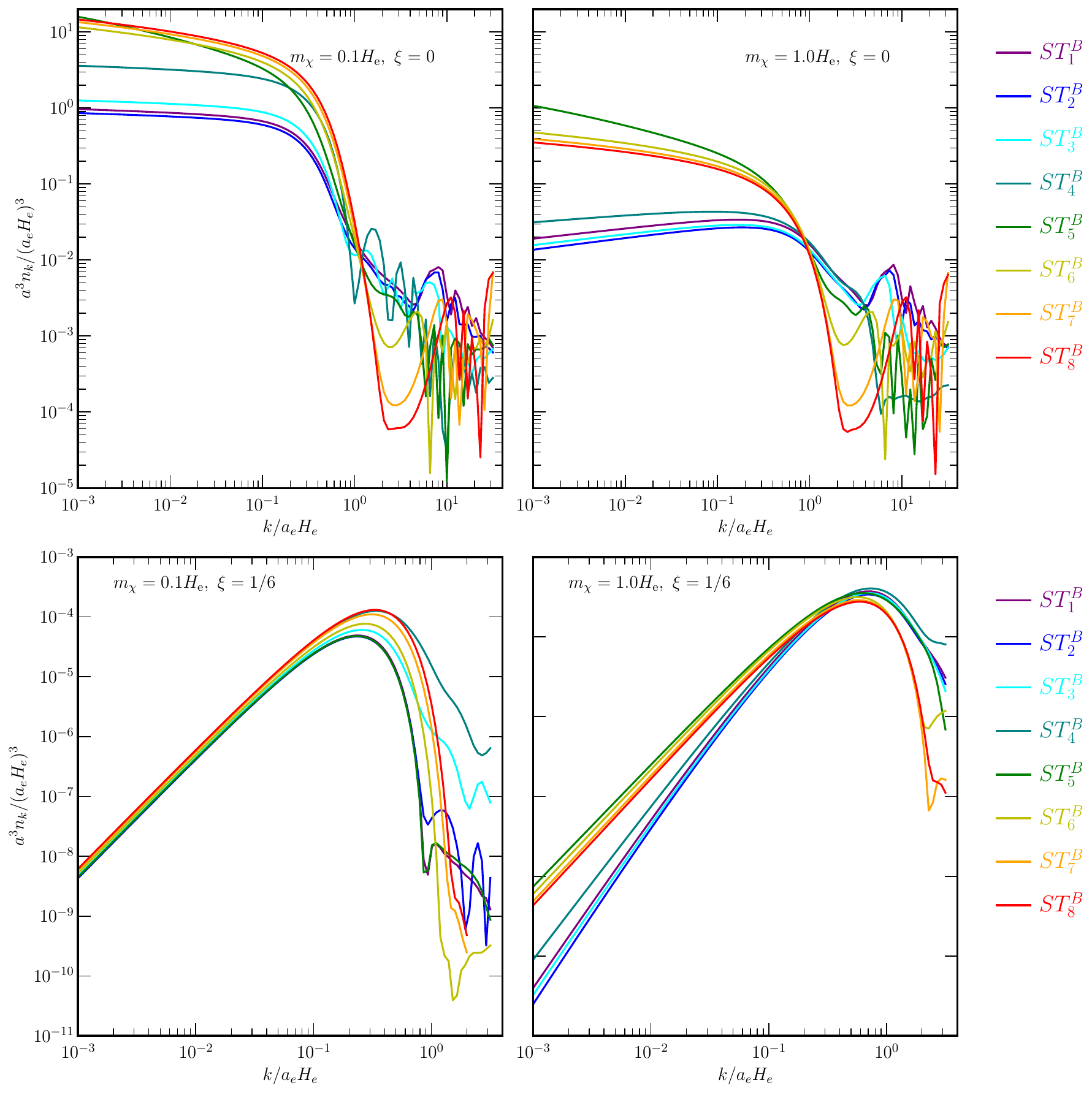}
\caption{Same as Fig.~\ref{fig:cgpp_STA} but for Series~B ($m_S/m_Q = 10$). The spectra show a wider spread across benchmarks, reflecting the stronger sensitivity of the lighter $\phi_Q$ field to the field-space curvature.}
\label{fig:cgpp_STB}
\end{figure}

This amplification has two consequences for the CGPP spectrum. For $\xi \neq 1/6$, the enhanced $R$ oscillations directly increase the nonadiabatic variation of $\omega_k^2$ through the $(1 - 6\xi)\,\partial_\eta^2 a/a$ coupling, producing more particles per oscillation cycle and extending the UV interference pattern to higher momenta. Each zero crossing of $R$ contributes a particle-production event~\cite{Basso:2022tpd}, and the denser spacing of zero crossings in the curved-geometry models (cf.\ the bottom panels of Figs.~\ref{fig:ricciSTA}--\ref{fig:ricciSTB}) translates into enhanced particle number densities. For $\xi = 1/6$, where the Ricci scalar drops out of $\omega_k^2$, the enhancement persists through a different channel: the more vigorous oscillation of $\dot\sigma^2$ modifies $H(t)$ via the Friedmann equation $\dot H = -\dot\sigma^2/(2M_{\rm Pl}^2)$, producing faster variations in $a(\eta)$ and hence stronger nonadiabatic evolution of the mass term $a^2 m_\chi^2$. The effect is weaker than in the $\xi = 0$ case, consistent with the reduced spread across benchmarks visible in the bottom panels of Figs.~\ref{fig:cgpp_STA} and~\ref{fig:cgpp_STB}.

A complementary perspective comes from the effective equation of state. For a single field oscillating in a quadratic potential, the time-averaged equation of state is $\langle w \rangle = 0$, and $R$ oscillates at frequency $2m$ around a slowly decaying mean $\langle R \rangle = 3H^2 \propto a^{-3}$, crossing zero twice per inflaton cycle. In the curved-geometry models, the geometric coupling transfers energy between the $\phi_S$ and $\phi_Q$ sectors during each oscillation cycle, modifying the instantaneous kinetic-to-potential-energy ratio and driving $w(t)$ away from zero more strongly than in the flat case. Since particle production is sensitive to the \emph{instantaneous} value of $R$, rather than only to its average, the larger excursions of $R$ in the curved-geometry models translate directly into larger particle yields.

The CGPP spectra for the sidetracked benchmarks are shown in Fig.~\ref{fig:cgpp_STA} (Series~A) and Fig.~\ref{fig:cgpp_STB} (Series~B), using the same spectator parameters as in the SQ case: $m_\chi/H_{\rm e} \in \{0.1, 1\}$ and $\xi \in \{0, 1/6\}$.

\emph{Minimally coupled case} ($\xi = 0$, top panels). The most striking feature is that decreasing $L_{\rm fs}$ (increasing field-space curvature) systematically enhances gravitational particle production. In the flat limit, ST$_1$ (purple) reproduces the pure Starobinsky spectrum and is the least efficient producer among the sidetracked benchmarks. As the field-space curvature increases through the stable regime (ST$_2$--ST$_5$), the IR plateau rises progressively, reaching roughly an order of magnitude above the flat limit for the most strongly curved stable runs. The sub-threshold runs (ST$_6$--ST$_8$) continue this trend, with ST$_8$ (red) producing up to an order of magnitude more particles than the flat limit in the IR.

This enhancement has a clear physical origin. Even in the stable regime, the geometric coupling between the two fields through the Christoffel symbols introduces additional oscillation modes that enrich the frequency content of $R(\eta)$, providing more zero crossings and hence more opportunities for nonadiabatic particle production. In the sub-threshold regime, the sidetracked trajectory involves correlated, high-frequency oscillations of both fields (cf.\ the bottom-right panels of Figs.~\ref{fig:ricciSTA} and~\ref{fig:ricciSTB}), which further amplify the post-inflationary oscillation amplitude of $R$ and extend particle production to higher $k$.

The UV interference structure also changes qualitatively with $L_{\rm fs}$. In the flat limit (ST$_1$), the UV tail is relatively smooth. As the curvature increases, the interference features become increasingly complex: the stable-regime runs (e.g., ST$_4$, ST$_5$) develop oscillatory UV tails from the multifrequency oscillations of $R$, while the sub-threshold runs (ST$_6$--ST$_8$) display rapid, dense interference patterns that extend to $k/k_{\rm e} \sim 10$--$30$, reflecting the high-frequency $R$ oscillations visible in Figs.~\ref{fig:ricciSTA} and~\ref{fig:ricciSTB}.

Comparing the light ($m_\chi = 0.1\,H_{\rm e}$, top left) and heavy ($m_\chi = H_{\rm e}$, top right) minimally coupled cases, the hierarchy across benchmarks is similar, but the heavy case shows somewhat tighter clustering of the IR plateau heights. This is because heavier particles are more strongly controlled by the common inflationary Hubble scale $H_{*}$, whereas differences in the post-inflationary oscillation pattern primarily affect the UV.

\emph{Conformally coupled case} ($\xi = 1/6$, bottom panels). The curvature coupling $(1 - 6\xi)\,\partial_\eta^2 a/a$ vanishes, and the spectra peak near $k \sim k_{\rm e}$, as in the SQ case. The overall normalization is suppressed by several orders of magnitude relative to the $\xi = 0$ case. The spread across benchmarks is reduced compared to the minimally coupled case, but a clear hierarchy persists: the sub-threshold runs still produce more particles than the flat limit, because the modified post-inflationary evolution of $a(\eta)$ on the sidetracked solution leads to faster variation of $a^2 m_\chi^2$ and hence stronger nonadiabatic particle production. For the heavy spectator ($m_\chi = H_{\rm e}$, bottom right), the spectra cluster more tightly around the peak, and the sub-threshold runs show only modest enhancement.

The spectra of Series~B (Fig.~\ref{fig:cgpp_STB}) display a qualitatively similar hierarchy to Series~A, but with a wider spread across benchmarks. This is because the lighter $\phi_Q$ field ($m_Q = m_S/10$) responds more strongly to the field-space curvature: the weaker restoring force $m^2_Q \phi_Q$ allows larger post-inflationary displacements, and the metric factor $e^{2\phi_Q/L_{\rm fs}}$ correspondingly amplifies the contribution of $\phi_S$ to $\dot{\sigma}^2$ and hence to $R$ [Eq. (\ref{eq:R_background})]. This is consistent with the field-space trajectories in Fig.~\ref{fig:STB_trajectories} and the Ricci scalar evolution in Fig.~\ref{fig:ricciSTB}, where the Series B runs display larger-amplitude, longer-timescale modulation of R than their Series A counterparts.

The sidetracked benchmarks demonstrate three distinct effects of field-space curvature on gravitational particle production:
\begin{enumerate}
\item \emph{Enhanced post-inflationary $R$ oscillations.} The geometric coupling between the inflaton fields, mediated by the curved field-space metric, introduces additional oscillation modes and increases the frequency content and amplitude of $R$ after inflation. This directly enhances particle production for $\xi \neq 1/6$ through the $(1 - 6\xi)\,\partial_\eta^2 a/a$ term in the mode equation~\eqref{eq:omega_k}.

\item \emph{Richer UV interference structure.} The multifrequency content of $R(\eta)$ in the curved field-space models produces complex interference patterns in the UV tail of the spectrum, in contrast to the smooth Starobinsky-like falloff or the regular staircase pattern of single-field quadratic inflation. These features encode detailed information about the post-inflationary field-space trajectory and could in principle serve as observational discriminants between single-field and multifield inflationary histories, although the relevant scales are typically superheavy and not directly accessible.

\item \emph{Systematic enhancement of the total abundance.} Across the benchmark set studied here, decreasing $L_{\rm fs}$ at fixed potential increases the total number density of gravitationally produced $\chi$ particles. For $\xi = 0$ and $m_\chi = 0.1\,H_{\rm e}$, the enhancement relative to the flat limit reaches approximately one order of magnitude for the deepest sub-threshold runs. This shifts the viable dark matter parameter space, i.e.\ the region in the $(m_\chi, T_{\rm RH})$ plane where $\Omega_\chi h^2 = 0.12$, toward lower $m_\chi$ or lower $T_{\rm RH}$ relative to the single-field Starobinsky case.
\end{enumerate}

\section{Dark Matter Abundance}
\label{sec:dm}

\subsection{Relic Abundance}
\label{subsec:relicabundance}
If the gravitationally produced $\chi$ particles are stable, they constitute a viable dark matter candidate. In this subsection we derive the present-day relic abundance $\Omega_\chi h^2$ in terms of the CGPP spectrum, following the treatment of Refs.~\cite{Kolb:2023ydq, Chung:1998zb}.

The comoving number density $n_\chi a^3$ is conserved after production, since $\chi$ interacts only gravitationally and backreaction on the inflaton condensate is negligible. For particles that are nonrelativistic today, the energy density and number density are related by $\rho_\chi(t_0) = m_\chi\,n_\chi(t_0)$. Writing
\begin{equation}
n_\chi(t_0) \;=\; 
\frac{a_0^3\,n_\chi(t_0)}{a_{\rm e}^3\,H_{\rm e}^3} 
\times \frac{a_{\rm e}^3\,H_{\rm e}^3}{a_0^3} \,,
\label{eq:n0_split}
\end{equation}
the first factor is a dimensionless number determined entirely by the CGPP calculation.\footnote{Thermal freeze-in from Standard Model initial states via $s$-channel graviton exchange, ${\rm SM}+{\rm SM}\to\chi+\chi$, contributes a density that depends on $m_\chi$ and $T_{\rm RH}$~\cite{Garny:2015sjg, Garny:2017kha, Tang:2017hvq, Mambrini:2021zpp, Clery:2021bwz}. In the parameter regime considered here, this contribution is always subdominant to the gravitational production we compute. The perturbative production from the inflaton condensate, ${\rm inflaton}+{\rm inflaton}\to\chi+\chi$~\cite{Ema:2018ucl, Mambrini:2021zpp}, is automatically captured by the Bogoliubov approach used in this work~\cite{Kaneta:2022gug}.} Since $a_0^3\,n_\chi(t_0) = a^3\,n_\chi(t)$ by comoving conservation, it equals
\begin{equation}
\frac{a_0^3\,n_\chi(t_0)}{a_{\rm e}^3\,H_{\rm e}^3} 
\;=\; \int d\log k\;\, \frac{a^3 \, n_k}{(a_{\rm e} H_{\rm e})^3}\,,
\label{eq:GPP_number}
\end{equation}
where we used Eq.~\eqref{eq:spectrum_cgpp}. This is precisely the quantity obtained by integrating the plotted spectra in Figs.~\ref{fig:cgpp_SQ}, \ref{fig:cgpp_STA}, and~\ref{fig:cgpp_STB}: the integrand is the $y$-axis and the integration variable is $\log[k/(a_{\rm e}H_{\rm e})]$.

The second factor in Eq.~\eqref{eq:n0_split} encodes the cosmological dilution from the end of inflation to today. We assume the inflaton oscillates coherently with an approximately matter-like equation of state ($\langle w \rangle \simeq 0$) from $t_{\rm e}$ until the reheating time $t_{\rm RH}$, at which the inflaton has fully decayed and the universe thermalizes at temperature $T_{\rm RH}$. During this phase, $\rho\,a^3 = \text{const}$, giving
\begin{equation}
\frac{a_{\rm e}^3}{a_{\rm RH}^3} \;=\; 
\frac{\rho_{\rm RH}}{\rho_{\rm e}} \;=\; 
\frac{\frac{\pi^2}{30}\,g_{*,{\rm RH}}\,T_{\rm RH}^4}
{3\,H_{\rm e}^2\,M_{\rm Pl}^2} \,,
\label{eq:ae_arh}
\end{equation}
where $\rho_{\rm e} = 3H_{\rm e}^2 M_{\rm Pl}^2$ from the Friedmann equation and $\rho_{\rm RH} = (\pi^2/30)\,g_{*,{\rm RH}}\,T_{\rm RH}^4$ is the energy density of the relativistic thermal bath at reheating. After reheating, entropy is conserved ($g_{*s}\,T^3 a^3 = \text{const}$), so
\begin{equation}
\frac{a_{\rm RH}^3}{a_0^3} \;=\; 
\frac{g_{*s,0}}{g_{*s,{\rm RH}}}\,
\frac{T_0^3}{T_{\rm RH}^3} \,,
\label{eq:arh_a0}
\end{equation}
where $g_{*s,0} \simeq 3.91$ is the present-day effective number of entropic degrees of freedom and $T_0 \simeq 2.725\,$K $\simeq 2.35 \times 10^{-13}\,$GeV is the CMB temperature today. Combining Eqs.~\eqref{eq:ae_arh} and~\eqref{eq:arh_a0} gives the redshift factor
\begin{equation}
\frac{a_{\rm e}^3\,H_{\rm e}^3}{a_0^3} \;=\; 
\frac{\pi^2\,g_{*,{\rm RH}}\,g_{*s,0}}
{90\,g_{*s,{\rm RH}}} \;\,
\frac{H_{\rm e}\,T_{\rm RH}\,T_0^3}{M_{\rm Pl}^2} \,.
\label{eq:redshift_factor}
\end{equation}

The relic abundance is expressed as $\Omega_\chi h^2 \equiv \rho_\chi(t_0)/\rho_{\rm crit} \times h^2$, where $h = H_0/(100\,\text{km}\,\text{s}^{-1}\,\text{Mpc}^{-1})$ parametrizes the present Hubble rate and $\rho_{\rm crit} = 3H_0^2 M_{\rm Pl}^2$ is the critical density. The combination $\rho_{\rm crit}/h^2 = 3(100\,\text{km}\,\text{s}^{-1}\,\text{Mpc}^{-1})^2 M_{\rm Pl}^2 \simeq 8.1 \times 10^{-47}\,\text{GeV}^4$ is independent of $H_0$. Substituting Eqs.~\eqref{eq:n0_split}--\eqref{eq:redshift_factor} into $\Omega_\chi h^2 = m_\chi\,n_\chi(t_0)/(\rho_{\rm crit}/h^2)$ and evaluating the numerical prefactor with $g_{*,{\rm RH}} = g_{*s,{\rm RH}} = 106.75$ for the Standard Model, the relic abundance takes the form
\begin{equation}
\frac{\Omega_\chi h^2}{0.12} 
\;\simeq\;
\left(\frac{m_\chi}{H_{\rm e}}\right)
\left(\frac{H_{\rm e}}{10^{12}\,\text{GeV}}\right)^{\!2}
\left(\frac{T_{\rm RH}}{10^{9}\,\text{GeV}}\right)
\left(\frac{1}{10^{-5}}\;
\frac{a_0^3\,n_\chi(t_0)}{a_{\rm e}^3\,H_{\rm e}^3}\right)\,,
\label{eq:Omega_scaling}
\end{equation}
where $\Omega_{\rm DM}h^2 \simeq 0.12$ is the observed dark matter relic abundance~\cite{Planck:2018jri}. Note that the dimensionless CGPP number $a_0^3 n_\chi(t_0)/(a_{\rm e}^3 H_{\rm e}^3)$ itself depends on $m_\chi/H_{\rm e}$ and $\xi$, and therefore introduces additional mass dependence beyond the explicit prefactor of $m_\chi/H_{\rm e}$.

Several features of Eq.~\eqref{eq:Omega_scaling} are worth noting. The abundance depends \emph{quadratically} on the inflationary Hubble scale $H_{\rm e}$: CGPP produces a comoving density that scales as $H_{\rm e}^3$, but the subsequent matter-dominated dilution during reheating removes one power via $\rho_{\rm e} \propto H_{\rm e}^2$, leaving a net $H_{\rm e}^2$ scaling (at fixed $m_\chi/H_{\rm e}$). The linear dependence on $T_{\rm RH}$ reflects the duration of the matter-dominated reheating phase: a higher reheating temperature means earlier thermalization, less dilution between $t_{\rm e}$ and $t_{\rm RH}$, and hence a larger surviving $\chi$ density. The multifield effects discussed in Secs.~\ref{subsec:cgpp_SQ}--\ref{subsec:cgpp_ST} modify $a_0^3 n_\chi(t_0)/(a_{\rm e}^3 H_{\rm e}^3)$ at fixed $m_\chi/H_{\rm e}$ and $\xi$: for example, the order-of-magnitude enhancement of this quantity in the sidetracked models relative to the flat Starobinsky limit (Sec.~\ref{subsec:cgpp_ST}) shifts the viable region in the $(m_\chi,\,T_{\rm RH})$ plane, allowing the correct relic abundance to be achieved at correspondingly lower values of $m_\chi$ or $T_{\rm RH}$.

For light spectators with $m_\chi \ll H_{\rm e}$ and minimal 
coupling ($\xi = 0$), the IR plateau in the spectrum 
$n_k/(a_{\rm e}H_{\rm e})^3$ is nearly flat (cf.\ top-left 
panels of Figs.~\ref{fig:cgpp_SQ}--\ref{fig:cgpp_STB}), 
and the integral~\eqref{eq:GPP_number} diverges logarithmically as $k \to 0$. A natural observable infrared cutoff is set by the present Hubble scale: modes with comoving wavenumber $k<k_0\equiv a_0H_0$ have wavelengths larger than the present horizon and do not contribute to the locally observable dark matter density. In the dimensionless units of the plots, this cutoff is
\begin{equation}
\frac{k_0}{a_{\rm e}\,H_{\rm e}} 
\;=\; \frac{a_0\,H_0}{a_{\rm e}\,H_{\rm e}} \,,
\label{eq:k_IR}
\end{equation}
which can be evaluated using 
Eqs.~\eqref{eq:ae_arh}--\eqref{eq:arh_a0}:
\begin{equation}
\ln\!\left(\frac{a_{\rm e}\,H_{\rm e}}{k_0}\right)
\;\equiv\; N_0 \;\simeq\; 
54.9 + \frac{1}{3}\,
\ln\!\left(\frac{H_{\rm e}}{10^{12}\,\text{GeV}}\right) 
+ \frac{1}{3}\,
\ln\!\left(\frac{T_{\rm RH}}{10^9\,\text{GeV}}\right) ,
\label{eq:N0}
\end{equation}
where we have used 
$H_0 = 67.4\,\text{km}\,\text{s}^{-1}\,\text{Mpc}^{-1}$ 
($h = 0.674$)~\cite{Planck:2018jri}. The quantity $N_0$ 
is the number of $e$-folds of expansion between the 
horizon exit of the mode $k_0$ during inflation and the 
end of inflation. 

For an approximately flat IR plateau of height $C \equiv a^3 n_k/(a_{\rm e}H_{\rm e})^3\big|_{k \ll k_{\rm e}}$, the integrated comoving number density~\eqref{eq:GPP_number} receives an IR contribution
\begin{equation}
\frac{a_0^3\,n_\chi(t_0)}{a_{\rm e}^3\,H_{\rm e}^3}
\bigg|_{\rm IR}
\;\simeq\; C \times N_0 \,,
\label{eq:IR_contribution}
\end{equation}
which dominates the total for $m_\chi \ll H_{\rm e}$ and $\xi = 0$, since the UV contribution from $k \gtrsim k_{\rm e}$ is confined to a narrow range in $\log k$. The logarithmic enhancement by $N_0 \sim 55$ means that the total abundance is set not by the peak of the spectrum but by the integrated weight of the nearly scale-invariant superhorizon modes accumulated over the entire observable inflationary history. This IR contribution is absent for $\xi = 1/6$ (conformal coupling), where the spectrum is steeply suppressed at $k \ll k_{\rm e}$, and for heavy spectators $m_\chi \gtrsim H_{\rm e}$, where the plateau height is exponentially suppressed by $\sim e^{-2\pi m_\chi/H_{*}}$.

\begin{figure}[t]
\centering
\includegraphics[width=0.5\textwidth]{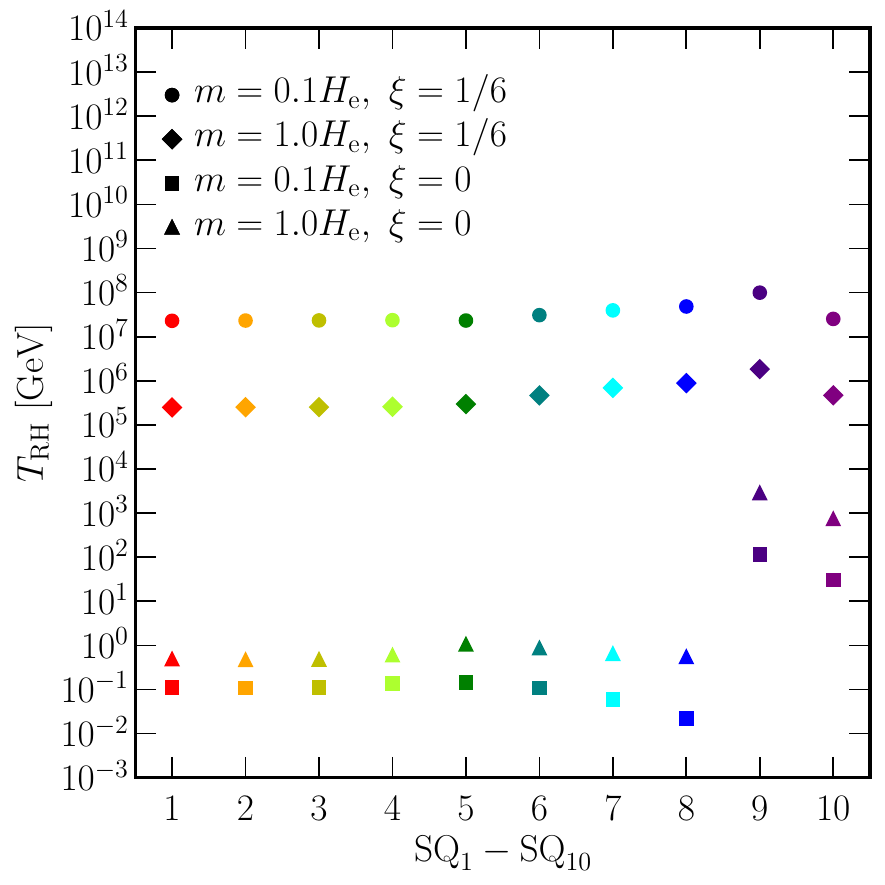}\\
\includegraphics[width=\textwidth]{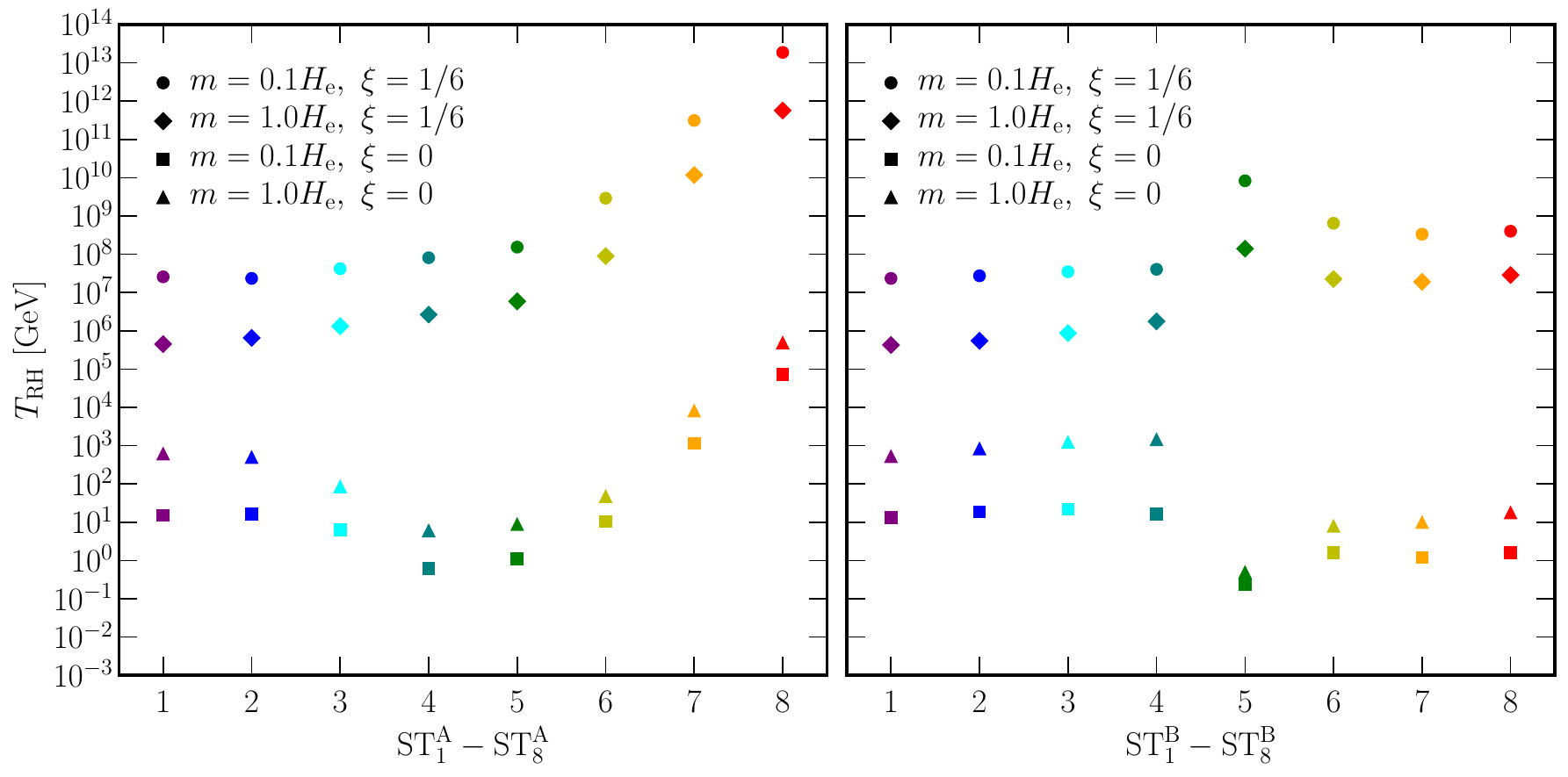}
\caption{Reheating temperature $T_{\rm RH}$ required for the gravitationally produced $\chi$ to constitute all of the dark matter ($\Omega_\chi h^2 = 0.12$), as a function of benchmark models, obtained from Eq.~\eqref{eq:Omega_scaling}.  Top: Starobinsky$+$quadratic benchmarks SQ$_1$--SQ$_{10}$ (Table~\ref{tab:SQ_runs}). Bottom left: sidetracked Series~A ($m_S/m_Q = 1.92$, Table~\ref{tab:ST_runs}). Bottom right: sidetracked Series~B ($m_S/m_Q = 10$). The color coding matches Figs.~\ref{fig:cgpp_SQ}--\ref{fig:cgpp_STB}. For $\xi = 0$, the required $T_{\rm RH}$ is very low ($\lesssim 10\,$GeV for the SQ benchmarks), reflecting the high efficiency of minimally coupled gravitational production. For $\xi = 1/6$, $T_{\rm RH}$ is four to six orders of magnitude higher. In the sidetracked benchmarks, the sub-threshold models (ST$^{A}_5$--ST$^{A}_8$) show a sharp increase in $T_{\rm RH}$ driven by the reduced inflationary energy scale $H_{\rm e}$ on the sidetracked attractor.}
\label{fig:dmplots}
\end{figure}

We use Eq.~\eqref{eq:Omega_scaling} to determine the 
reheating temperature $T_{\rm RH}$ required for the 
gravitationally produced $\chi$ to saturate the observed 
dark matter abundance, $\Omega_\chi h^2 = 0.12$, for each 
benchmark model. The results are shown in 
Fig.~\ref{fig:dmplots} for four combinations of spectator 
parameters: $m_\chi/H_{\rm e} \in \{0.1,\,1\}$ and 
$\xi \in \{0,\,1/6\}$.

\paragraph{Starobinsky$+$quadratic benchmarks.}
The top panel of Fig.~\ref{fig:dmplots} displays 
$T_{\rm RH}$ for the SQ benchmarks. For minimal coupling 
($\xi = 0$, squares/triangles), the required reheating temperatures 
are remarkably low: $T_{\rm RH} \sim 0.01$--$10\,$GeV, 
reflecting the high efficiency of gravitational production 
when the curvature coupling 
$(1-6\xi)\,\partial_\eta^2 a/a$ is active. The ordering 
follows the CGPP spectra of Fig.~\ref{fig:cgpp_SQ}: 
the quadratic-dominated runs (SQ$_1$--SQ$_3$) produce 
the most particles and therefore require the lowest 
$T_{\rm RH}$ to avoid overproduction, while the 
Starobinsky-dominated runs (SQ$_9$--SQ$_{10}$) cluster 
at higher $T_{\rm RH}$. The spread across the full SQ 
scan is roughly one order of magnitude for $\xi = 0$, 
reflecting the modest variation of the IR plateau height 
across benchmarks.

For conformal coupling ($\xi = 1/6$, circles/diamonds), the 
required $T_{\rm RH}$ is four to six orders of magnitude 
higher, in the range $10^4$--$10^8\,$GeV. This 
dramatic shift reflects the suppression of particle 
production when the curvature coupling is absent: without 
the tachyonic superhorizon amplification, the CGPP number 
density is orders of magnitude smaller, and a 
correspondingly higher $T_{\rm RH}$ (i.e., less 
matter-dominated dilution) is needed to compensate. The 
spread across benchmarks is also wider for $\xi = 1/6$ 
than for $\xi = 0$, because the conformal spectrum is 
more sensitive to the post-inflationary oscillation vigor, 
which varies strongly between the quadratic and 
Starobinsky limits.

Within each coupling, the heavier spectator 
($m_\chi = H_{\rm e}$) requires a higher 
$T_{\rm RH}$ than the lighter one 
($m_\chi = 0.1\,H_{\rm e}$) for $\xi = 0$, because the 
CGPP number density decreases with increasing 
$m_\chi/H_{\rm e}$ due to the Boltzmann-like 
suppression $\sim e^{-2\pi m_\chi/H}$ of the superhorizon 
plateau. For $\xi = 1/6$, the ordering is reversed: heavier particles are produced more efficiently, and the explicit factor of $m_\chi/H_{\rm e}$ in Eq.~\eqref{eq:Omega_scaling} further increases the abundance, so the required $T_{\rm RH}$ is lower for $m_\chi = H_{\rm e}$.

\paragraph{Sidetracked benchmarks.}
The bottom panels of Fig.~\ref{fig:dmplots} show the 
corresponding results for Series~A (left) and Series~B 
(right). In the stable regime 
(ST$_1$--ST$_5$), the required $T_{\rm RH}$ 
decreases monotonically as $L_{\rm fs}$ decreases 
(increasing field-space curvature), consistent with the 
systematic enhancement of the CGPP number density 
discussed in Sec.~\ref{subsec:cgpp_ST}. For $\xi = 0$, 
the stable-regime models span roughly one order of 
magnitude in $T_{\rm RH}$, with the most strongly 
curved models (ST$_4$, ST$_5$) requiring $T_{\rm RH}$ a 
factor of a few below the flat limit.

The threshold and sub-threshold models 
(ST$_6$--ST$_8$) exhibit qualitatively different behavior, 
driven by the dramatic reduction of the inflationary 
energy scale on the sidetracked attractor. For 
ST$^{\rm A}_8$, the Starobinsky-sector mass drops to 
$m_S \simeq 3 \times 10^{-7}\,M_{\rm Pl}$, giving 
$H_{\rm e} \simeq 3.7 \times 10^{11}\,$GeV---more than 
an order of magnitude below the flat-limit value 
$H_{\rm e} \simeq 1.2 \times 10^{13}\,$GeV. Since the 
relic abundance scales as $H_{\rm e}^2$ at fixed 
$m_\chi/H_{\rm e}$ [Eq.~\eqref{eq:Omega_scaling}], this 
suppresses the $\chi$ density by a factor of 
$(H_{{\rm e},\,{\rm ST}_8}/H_{{\rm e},\,{\rm ST}_1})^2 
\sim 10^{-3}$, which overwhelms the order-of-magnitude 
enhancement of the CGPP number density from the more 
vigorous post-inflationary oscillations. The net effect 
is a sharp \emph{increase} in the required $T_{\rm RH}$ 
for the sub-threshold models, particularly visible for 
$\xi = 1/6$ where $T_{\rm RH}$ reaches 
$10^{12}$--$10^{13}\,$GeV for ST$^{\rm A}_8$. This 
illustrates the competing roles of field-space curvature 
in gravitational dark matter production: while negative 
curvature enhances the CGPP efficiency per Hubble volume, 
it simultaneously lowers the inflationary energy scale 
on the sidetracked attractor, and the latter effect 
dominates.

Series~B (bottom right) shows the same qualitative 
structure as Series~A, with the lighter $\phi_Q$ field 
amplifying both effects: the stable-regime enhancement 
is somewhat stronger, and the sub-threshold suppression 
from the reduced $H_{\rm e}$ is equally severe. The 
net result is that the viable dark matter parameter 
space depends sensitively on the interplay between 
field-space curvature, the inflationary energy scale, 
and the reheating temperature, making gravitational 
particle production a probe of the geometry of the 
inflaton field space.

\subsection{Isocurvature Constraints}
\label{subsec:iso}
We next impose the dark matter isocurvature constraint on the gravitationally produced $\chi$ abundance. This should be distinguished from the entropic fluctuation orthogonal to the multifield inflationary trajectory, which controls the usual adiabatic--isocurvature mixing during inflation. In what follows, we refer to the former as the \emph{dark matter isocurvature} (associated with the spectator $\chi$ abundance) and to the latter as the \emph{inflationary entropic mode}

A gravitationally produced spectator field $\chi$ that 
constitutes dark matter generically carries isocurvature 
perturbations, since its density fluctuations are seeded 
by quantum fluctuations during inflation and need not 
coincide with the adiabatic perturbations in the radiation 
sector~\cite{Chung:2004nh, Chung:2011hv, Kolb:2022eyn, Garcia:2022vwm, Garcia:2023awt}. 
These dark-matter isocurvature perturbations are 
constrained by Planck to satisfy
\begin{equation}
\beta_{\rm iso}
\;\equiv\;
\frac{\mathcal P_{\rm iso}}
{\mathcal P_{\mathcal R}+\mathcal P_{\rm iso}}
\;< \;0.038
\label{eq:beta_iso}
\end{equation}
at 95\% CL~\cite{Planck:2018jri}.

For a light minimally coupled scalar ($\xi=0$, 
$m_\chi \ll H_{*}$), the field acquires nearly 
scale-invariant fluctuations during inflation with 
amplitude
\begin{equation}
\delta\chi_* \;\sim\; \frac{H_{*}}{2\pi} \,.
\end{equation}
The resulting isocurvature perturbation is controlled by 
how sensitively the local dark matter energy density 
depends on the superhorizon field value at horizon exit. 
Denoting the final yield by 
$Y_\chi \equiv n_\chi/s$, one has schematically
\begin{equation}
S_\chi
\;\equiv\;
\delta \ln Y_\chi
\;\simeq\;
\frac{\partial \ln Y_\chi}{\partial \chi_*}\,
\delta\chi_* \,,
\label{eq:Schi_def}
\end{equation}
where the dependence on $\chi_*$ enters through 
the contribution of the superhorizon condensate to the 
local energy density. The isocurvature power spectrum 
then scales as
\begin{equation}
\mathcal P_{\rm iso}
\;\sim\;
\left(\frac{\Omega_\chi}{\Omega_{\rm DM}}\right)^2
\left(\frac{\partial \ln Y_\chi}{\partial \chi_*}
\right)^2
\left(\frac{H_{*}}{2\pi}\right)^2 .
\label{eq:P_iso}
\end{equation}
For light minimally coupled spectators, the inflationary 
fluctuations can easily generate an unacceptably large 
isocurvature signal if $\chi$ constitutes all of the dark 
matter. In practice, this strongly disfavors the 
light-mass regime $m_\chi \ll H_{*}$ for $\xi=0$, 
including the representative case 
$m_\chi = 0.1\,H_{\rm e}$ considered in our benchmark 
study, unless the isocurvature contribution is diluted or 
otherwise 
suppressed~\cite{Chung:2004nh, Chung:2011hv, Kolb:2022eyn}. 
By contrast, the heavier minimally coupled case 
$m_\chi \sim H_{\rm e}$ is expected to be less 
constrained, although its precise viability depends on 
the detailed mapping between the inflationary spectator 
fluctuations and the final abundance.

The constraint is significantly relaxed for conformal 
coupling ($\xi=1/6$). In this case, particle production 
is driven by the mass term $a^2m_\chi^2$ rather than by 
the curvature coupling, and the final abundance is much 
less sensitive to inflationary superhorizon fluctuations 
of $\chi$: production is concentrated around the end of 
inflation and during the post-inflationary oscillatory 
phase, where the background nonadiabaticity is dominated 
by the evolution of the inflaton 
sector~\cite{Kolb:2022eyn}. The $\xi=1/6$ benchmarks in 
Fig.~\ref{fig:dmplots} are therefore expected to be much 
less constrained by isocurvature, further motivating the 
conformally coupled scalar, and by extension fermionic, 
dark matter scenario.

We emphasize that the isocurvature bound applies most 
directly to the case in which gravitationally produced 
$\chi$ constitutes \emph{all} of the dark matter. If 
$\chi$ is only a subdominant component, the isocurvature 
amplitude is suppressed by 
$(\Omega_\chi/\Omega_{\rm DM})^2$ 
[Eq.~\eqref{eq:P_iso}], and correspondingly lighter 
masses become viable. A dedicated analysis of 
isocurvature in the multifield backgrounds considered 
here is left for future work.

\section{Discussion and Conclusions}
\label{sec:conclusions}

In this paper we have studied cosmological gravitational particle production (CGPP) of a spectator scalar field $\chi$ in two-field inflationary models with both flat and curved field-space geometries. Working throughout with the Starobinsky$+$quadratic potential~\eqref{eq:V_SQ}, we have systematically disentangled two distinct sources of modification to the CGPP spectrum relative to single-field inflation: the effect of initial conditions in field space at fixed flat geometry (Sec.~\ref{subsec:starobinsky_quadratic}), and the effect of field-space curvature at fixed potential and initial conditions (Sec.~\ref{subsec:sidetracked}). The spectator $\chi$ couples only to gravity and serves as a dark matter candidate; for conformal coupling $\xi = 1/6$, its production spectrum provides a useful proxy for that of a massive Dirac fermion, up to the appropriate spin degeneracy factor.

Our main results are as follows. We constructed two comprehensive sets of benchmark models. The Starobinsky$+$quadratic scan (SQ$_1$--SQ$_{10}$, Table~\ref{tab:SQ_runs}) interpolates continuously between the pure quadratic and pure Starobinsky limits on a flat field space, with only the most Starobinsky-dominated runs (SQ$_9$, SQ$_{10}$) satisfying the joint Planck$+$BK18 constraints. The sidetracked scan (Table~\ref{tab:ST_runs}) introduces negative field-space curvature through a hyperbolic metric parametrized by the curvature $L_{\rm fs}$, at two mass ratios ($m_S/m_Q = 1.92$ and $10$). We identified three distinct regimes as a function of $L_{\rm fs}/L_{\rm crit}$: a stable regime with perturbative corrections to Starobinsky inflation, a forbidden band near the destabilization threshold where $n_s$ falls outside the observationally allowed range, often with $n_s \geq 1$, and a sub-threshold sidetracked regime that can remain observationally viable while yielding dramatically reduced tensor-to-scalar ratios, $r \sim 10^{-7}$, which for our benchmark parameters is nevertheless excluded by its blue scalar tilt.

The spectator field $\chi$ does not couple directly to the inflaton field-space metric $G_{IJ}$; the effect of multifield dynamics on CGPP enters only through the spacetime quantities $a(\eta)$ and $R(\eta)$ that appear in the mode equation~\eqref{eq:mode_eq_cgpp}. In the flat Starobinsky$+$quadratic models, the two fields oscillate independently to leading order after inflation, and the Ricci scalar inherits a beating pattern from the superposition of oscillations at frequencies primarily set by $2m_S$ and $2m_Q$ (Fig.~\ref{fig:ricciSQ}). In the sidetracked models, the Christoffel symbols of the curved field-space metric act as a geometric pump that sources $\phi_Q$ oscillations and, through the metric factor $e^{2\phi_Q/L_{\rm fs}}$ in the kinetic term, amplifies the contribution of $\phi_S$ to $\dot\sigma^2$ and hence to $R$. This produces high-frequency, large-amplitude oscillations of $R$ after inflation (Figs.~\ref{fig:ricciSTA}--\ref{fig:ricciSTB}) that are qualitatively distinct from either the gentle Starobinsky oscillation or the regular quadratic oscillation of the single-field limits.

For minimal coupling ($\xi = 0$), the CGPP spectrum exhibits a nearly scale-invariant IR plateau set primarily by the inflationary Hubble scale, together with a UV region sensitive to the post-inflationary oscillation pattern. The IR plateau height varies by a factor of three to five across the flat SQ benchmarks (Fig.~\ref{fig:cgpp_SQ}), with the quadratic-dominated runs producing the most particles. In the sidetracked models (Figs.~\ref{fig:cgpp_STA}--\ref{fig:cgpp_STB}), decreasing the field-space curvature $L_{\rm fs}$ systematically enhances gravitational particle production: the CGPP number density increases by up to an order of magnitude relative to the flat Starobinsky limit, driven by the geometry-enhanced oscillations of $R$. The UV interference structure also becomes increasingly complex, encoding detailed information about the post-inflationary field-space trajectory. For conformal coupling ($\xi = 1/6$), the IR plateau is eliminated and production is suppressed by several orders of magnitude, but the hierarchy across benchmarks persists through the modified evolution of $a(\eta)$.

The relic abundance [Eq.~\eqref{eq:Omega_scaling}] depends quadratically on the inflationary Hubble scale $H_{\rm e}$, linearly on the reheating temperature $T_{\rm RH}$, and on the dimensionless CGPP number density that encodes the full dependence on $m_\chi/H_{\rm e}$, $\xi$, and the inflationary background. For the sidetracked models, the enhanced CGPP efficiency competes with the reduced inflationary energy scale on the sidetracked attractor: the sub-threshold model ST$^{\rm A}_8$ has $H_{\rm e}$ more than an order of magnitude below the flat-limit value, and the resulting $H_{\rm e}^2$ suppression dominates over the order-of-magnitude enhancement of the CGPP number density (Fig.~\ref{fig:dmplots}). This interplay between geometry-enhanced production efficiency and geometry-suppressed inflationary energy scale is a distinctive feature of gravitational dark matter production in models with curved field spaces.

The Planck isocurvature bound~\cite{Planck:2018jri} strongly disfavors the light-mass regime $m_\chi \ll H_{*}$ with $\xi = 0$ as a scenario in which gravitationally produced $\chi$ constitutes all of the dark matter, because inflationary superhorizon fluctuations of the spectator can induce an isocurvature signal that is too large. This constraint is significantly relaxed for conformal coupling $\xi = 1/6$, where production is driven by the mass term rather than by the curvature coupling. Since the $\xi = 1/6$ scalar mode equation is closely analogous to that of a massive Dirac fermion, the conformally coupled spectra presented here also provide a useful proxy for purely gravitational fermionic dark matter production.

Several directions merit further investigation. First, a more detailed treatment of reheating dynamics---including the decay channels of the multifield inflaton condensate and the resulting effective equation of state---would refine the $T_{\rm RH}$ predictions of Fig.~\ref{fig:dmplots}. Second, the isocurvature analysis could be extended using a full treatment of the spectator and inflaton perturbations in the multifield backgrounds considered here. Third, the framework developed here generalizes naturally to other rapid-turn attractors~\cite{Brown:2017osf, Bjorkmo:2019qno, Aragam:2021scu} and to higher-spin spectators, including massive vectors and spin-2 particles produced on multifield backgrounds. Finally, the sensitivity of the CGPP spectrum to the post-inflationary trajectory suggests that gravitational particle production, if the dark matter mass and abundance can be independently constrained, could serve as an indirect probe of the geometry of the inflaton field space.

\acknowledgments
We thank Marcos A. G. Garc\'ia, Andrew Long, Evan McDonough, and Keith Olive for useful discussions. The work of E.W.K., S.V., and J.W. was supported by the Kavli Institute for Cosmological Physics at the University of Chicago.

\addcontentsline{toc}{section}{References}
\bibliographystyle{utphys}
\bibliography{references}

\end{document}